\documentclass[letter]{article}
\usepackage{jheppub}
\usepackage{amsmath,amsfonts,physics,relsize,slashed}
\usepackage[capitalize]{cleveref}

\usepackage{graphicx}% Include figure files
\usepackage{dcolumn}% Align table columns on decimal point
\usepackage{bm}% bold math
\usepackage{hyperref}% add hypertext capabilities
\usepackage[mathlines]{lineno}% Enable numbering of text and display math
\usepackage[caption=false]{subfig}
\usepackage{diagbox}
\usepackage{tabularx}
\usepackage{xcolor}

\usepackage{floatrow}
\usepackage{comment}

\usepackage[rightcaption]{sidecap}
\sidecaptionvpos{figure}{c}
 
\newfloatcommand{capbtabbox}{table}[][\FBwidth]

\usepackage{enumitem}

\usepackage{xspace}

\renewcommand{\vec}[1]{\mathbf{#1}}

\newcommand{\isotope}[3]{${}^{#2}_{#1}\mathrm{#3}$}

\usepackage{color}

\definecolor{ColorSun1}{rgb}{1.0,0.75,0.11}
\definecolor{ColorSun2}{rgb}{0.72,0.09,0.01}

\title{Solar reflection of dark matter with dark-photon mediators
}

\author[a]{Timon Emken,}
\author[b]{Rouven Essig,}
\author[b]{Hailin Xu}

\affiliation[a]{Oskar Klein Centre, Department of Physics, Stockholm University, Stockholm SE-10691, Sweden}
\affiliation[b]{C.N. Yang Institute for Theoretical Physics, Stony Brook University, Stony Brook, NY 11794}

\emailAdd{timon.emken@gmail.com}
\emailAdd{rouven.essig@stonybrook.edu}
\emailAdd{hailin.xu@stonybrook.edu}

\preprint{YITP-SB-2024-07}

\abstract{
We consider the scattering of low-mass halo dark-matter particles in the hot plasma of the Sun, focusing on dark matter that interact with ordinary matter through a dark-photon mediator.  The resulting ``solar-reflected'' dark matter (SRDM) component contains high-velocity particles, which significantly extend the sensitivity of terrestrial direct-detection experiments to sub-MeV dark-matter masses.  We use a detailed Monte-Carlo simulation to model the propagation and scattering of dark-matter particles in the Sun, including thermal effects, with special emphasis on ultralight dark-photon mediators. We study the properties of the SRDM flux, obtain exclusion limits from various direct-detection experiments, and provide projections for future experiments, focusing especially on those with silicon and xenon targets.  We find that proposed future experiments with xenon and silicon targets can probe the entire ``freeze-in benchmark,'' in which dark matter is coupled to an ultralight dark photon, including dark-matter masses as low as $\mathcal{O}$(keV).  Our simulations and SRDM fluxes are publicly available.
}

\begin{document}
\maketitle
%\flushbottom

%\pagebreak 

%%%%%%%%%%%%%%%%%%%%%%%%%%%%%%%%%%%%%%%%%%%%%%%%%%%%%%%%%%%%%%%
\section{Introduction}
\label{sec: introduction}

The search for dark matter (DM) particles with masses below the proton has seen tremendous progress over the past few years~\cite{Essig:2022dfa}.  In particular, searches for DM-electron scattering~\cite{Essig:2011nj} and DM-nucleus scattering with the Migdal effect~\cite{Ibe:2017yqa} have begun probing DM down to $\sim$MeV-masses using noble liquid and semiconductor targets
~\cite{Angle:2011th,Essig:2012yx,Aprile:2016wwo,DAMIC:2016lrs,Essig:2017kqs,Crisler:2018gci,Agnes:2018oej,Agnese:2018col,Aguilar-Arevalo:2019wdi,SENSEI:2019ibb,XENON:2019gfn,Blanco:2019lrf,SENSEI:2020dpa,Amaral:2020ryn,XENON:2021qze,PandaX-II:2021nsg,Arnaud:2020svb,DAMIC-M:2023gxo,SENSEI:2023zdf}. To probe non-relativistic DM particles in our halo that have even lower masses (down to $\sim$keV) requires more ambitious detector technologies~\cite{Hochberg:2015pha, Hochberg:2015fth, Hochberg:2016ntt, Hochberg:2017wce, Knapen:2016cue, Knapen:2017ekk}.  However, if the halo DM interacts with our detector, some small fraction of it can also be accelerated.  In particular, DM particles can be accelerated by scattering off cosmic-rays in the galaxy~\cite{Bringmann:2018cvk, Cappiello:2018hsu, Ema:2018bih, Cappiello:2019qsw, Dent:2019krz,Bondarenko:2019vrb, Wang:2019jtk,Dent:2020syp} or in blazars~\cite{Wang:2021jic}, or through their scattering in the Sun~\cite{An:2017ojc,Emken:2017hnp,Emken:2019hgy,Chen:2020gcl,Emken:2021lgc,An:2021qdl,Zhang:2023xwv,Liang:2023ira}. 
The latter is the mechanism that probes the lowest interaction strengths and is the focus of this paper.    

The idea of solar-reflected and accelerated DM was first proposed in~\cite{An:2017ojc, Emken:2017hnp}. As the solar system is moving with respect to the galaxy, the halo DM particles enter the gravitational well of the Sun and pass through the solar plasma. There is some chance that DM particles scatter with an energetic electron or nucleus in the solar medium, which can significantly accelerate the DM particle compared to its original speed, even well above the galactic escape velocity. It results in an energetic flux of ``solar reflected DM'' (SRDM) that is emitted from the Sun. Searches for this SRDM flux enables terrestrial DM detectors to extend their sensitivity to lower DM masses. SRDM also features a rich phenomenology, such as an anisotropy of the SRDM flux and modulation of the signal rate. Solar reflection does not require additional model assumptions, as the interaction that guarantees that the DM particles scatter in the Sun are also needed to probe them with our detectors on Earth.  Therefore, the SRDM flux is an intrinsic feature of any DM particle scenario with sizable DM-matter interactions.

Ref.~\cite{Emken:2021lgc} used a Monte-Carlo (MC) simulation to calculate the SRDM flux and derives direct-detection bounds on DM that interacts with heavy mediators.  We extend this work to DM interactions mediated by an ultralight dark photon,\footnote{The solar-reflected DM flux for ultralight dark-photon mediators was first calculated in~\cite{An:2021qdl}; our work corrects two errors found in version~1 of~\cite{An:2021qdl}. Our results now agree with their update in version 2 (see comparison in \cref{app: xenon form factors}), despite both groups using independent simulation codes and analyses.} but also revisit the constraints and predictions for heavy dark-photon mediators.  In particular, we include now in-medium effects for the dark-photon interaction in the solar plasma and use an improved understanding of how boosted DM particles interact in silicon, taking into account the effect from collective excitations such as the plasmon~\cite{Essig:2024ebk}.  
 
The rest of the paper is organized as follows.  In \cref{sec: solarreflection}, we first review the solar reflection of light DM in general and independent of the specific DM model, before reviewing dark-photon mediators and discussing the direct-detection scattering rates. In \cref{sec: montecarlosimulations}, we introduce the new version of our MC tool Dark Matter Simulation Code for Underground Scatterings -- Sun Edition (\texttt{DaMaSCUS-SUN}~\cite{Emken2021}), which is expanded to incorporate in-medium effects and new calculations of scattering rates for DM with light mediators.\footnote{We make our simulation results for the SRDM fluxes publicly available at \url{https://github.com/hlxuhep/Solar-Reflected-Dark-Matter-Flux}.} In \cref{sec: results}, we show our MC results for the SRDM flux and spectrum, direct-detection limits and projections from solar reflection, and the annual modulation of the SRDM scattering rates.  We comment on the possibility of searching for a directional-detection signal of the SRDM flux with diurnal modulation. We give our conclusions in \cref{sec: conclusions}. We also have several appendices. In~\cref{app: inmedium}, we revisit the calculation of the in-medium effects of DM scattering in the solar plasma with a dark-photon mediator~\cite{DeRocco:2022rze}. We present details of our MC simulation in~\cref{app: simulation details}. In \cref{app: xenon form factors} and \cref{app: silicon form factors}, we detail the xenon and silicon form factors, respectively, and compare how different calculations of the form factors impact the results. 
%%%%%%%%%%%%%%%%%%%%%%%%%%%%%%%%%%%%%%%%%%%%%%%%%%%%%%%%%%%%%%%
\section{Solar reflection of dark matter}\label{sec: solarreflection}
In this section, we begin by reviewing the solar reflection of DM. We first discuss the properties of halo DM particles, and then discuss how these particles fall into the Sun's gravitational well. We then describe in general the scattering of DM with particles in the Sun, before focusing on models with a dark-photon mediator for which we account for in-medium effects. Finally, we discuss the scattering of solar-reflected DM particles in direct-detection experiments with noble-liquid and semiconductor targets.

\subsection{Dark matter in the galactic halo}
We assume the standard halo model (SHM) for the distribution of DM particles in the galactic halo as it is also the standard choice for the analysis of direct-detection experiments. In this model, the local DM particles have a constant mass density $\rho\approx 0.3\;\text{GeV\;cm}^{-3}$~\cite{Baxter:2021pqo} and follow a truncated Maxwell-Boltzmann velocity distribution,
\begin{equation}\label{eq: SHM velocity distribution}
    f_{\text{halo}}(\vec{v})=\frac{1}{N_{\text{esc}}\pi^{3/2}v_0^3}\exp\left(-\frac{\vec{v}^2}{v_0^2}\right)\Theta(v_{\text{gal}}-\left|\vec{v}\right|)\,,
\end{equation}
where the normalization constant is given by
\begin{equation}
    N_{\text{esc}}\equiv \erf\left(\frac{v_{\text{gal}}}{v_0}\right)-\frac{2}{\sqrt{\pi}}\frac{v_{\text{gal}}}{v_0}\exp\left(-\frac{v_{\text{gal}}^2}{v_0^2}\right)\,.
\end{equation}
The galactic escape velocity is $v_{\text{gal}}\approx544\;\text{km s}^{-1}$, and the velocity dispersion is $v_0\approx220\;\text{km s}^{-1}$~\cite{Strigari:2012acq}.

The DM velocity distribution in a direct-detection experiment with velocity $\vec{v}_{\text{obs}}$ with respect to the galactic rest frame is
\begin{equation}
    f_{\chi,\text{lab}}(\vec{v})=f_{\text{halo}}(\vec{v}+\vec{v}_{\text{obs}})\,,
\end{equation}
where $\vec{v}$ is the velocity of the DM particle in the lab frame. Hence the maximum velocity of a DM particle entering the detector is $v_{\text{max}}=v_{\text{gal}}+v_{\text{obs}}$. In an elastic DM-nucleus scattering event, the maximum nuclear recoil energy is $E_R^{\text{max}}=\frac{2\mu_{\chi N}^2}{m_N}v_{\text{max}}^2$, where $\mu_{\chi N}$ is the reduced mass of the DM mass $m_\chi$ and the mass of the nucleus $m_N$. The maximum recoil energy $E_R^{\text{max}}$ must exceed the detection threshold $E_{\text{thr}}$ so that the DM particle can possibly trigger the detector, which sets the lowest mass that this detector can reach,
\begin{equation}\label{nucleusdetectorreach}
    m_\chi\gtrsim\frac{m_N}{\sqrt{\frac{2m_N}{E_{\text{thr}}}v_{\text{max}}}-1}\,.
\end{equation}
For DM-electron scattering, the maximum recoil energy of the electron is $\Delta E_e\simeq\frac{1}{2}m_\chi v_\text{max}^2$, which is significantly larger than for elastic DM-nucleus scattering, at least for sub-GeV DM.  The minimum mass for DM to be detected is then
\begin{equation}\label{electrondetectorreach}
    m_\chi\gtrsim\frac{2E_{\text{gap}}}{v_\text{max}^2}\,.
\end{equation}
The SRDM component has a significantly larger $v_{\rm max}$ than halo DM, allowing one to probe lower DM masses. 

\subsection{Halo dark matter falling into the Sun's gravitational well}
For halo DM particles asymptotically far from the Sun, the distribution of their initial velocity $\vec{u}$ in the solar rest frame follows
\begin{equation}
    f_{\chi}(\vec{u})=f_{\text{halo}}(\vec{u}+\vec{v}_\odot)\,,
\end{equation}
where $\vec{v}_\odot$ is the Sun's velocity in the galactic rest frame, and $f_{\text{halo}}$ is the halo DM distribution from \cref{eq: SHM velocity distribution}. The initial velocity $\vec{u}$ specifies the direction of the incoming DM particle at large distances from the Sun. The maximum impact parameter for the DM particle with initial velocity $\vec{u}$ touching the Sun is specified by energy conservation and angular-momentum conservation,
\begin{equation}
    \begin{split}
        b_{\text{max}}\equiv\frac{\sqrt{u^2+v_{\text{esc}}(R_\odot)^2}R_{\odot}}{u}=\sqrt{1+\frac{v_{\text{esc}}(R_\odot)^2}{u^2}}R_\odot\,,
    \end{split}
\end{equation}
where $v_{\text{esc}}(R_\odot)\approx 618\;\text{km s}^{-1}$. Before the DM particle enters the Sun, it travels along a hyperbolic Kepler orbit. The differential rate $d\Gamma$ of DM falling into the Sun with asymptotic velocity $\vec{u}$ is
\begin{equation}\label{eq: diff_entering_rate}
    \begin{split}
        \dd\Gamma=n_\chi f_\chi(\vec{u}) u\pi b_{\text{max}}^2\dd^3\vec{u}=n_\chi\pi R_\odot^2f_\chi(\vec{u})\left(u+\frac{v_{\text{esc}}(R_\odot)^2}{u}\right)\dd^3\vec{u}\,.
    \end{split}
\end{equation}
We use \cref{eq: diff_entering_rate} to sample the initial velocity of incoming halo DM particles, which correctly incorporates the gravitational lensing effect. The initial position far from the Sun is sampled from a homogeneous distribution within the two dimensional plane with impact parameter $b<b_{\text{max}}$. Details of our Monte Carlo simulations are given in \cref{sec: montecarlosimulations}. We can estimate the total rate~$\Gamma$ of DM particle falling into the Sun to be
\begin{subequations}
\label{eq: infalling rate}
\begin{align}
        \Gamma&=n_\chi\pi R_\odot^2\int \dd^3\vec{u}\,f_\chi(\vec{u})\left(u+\frac{v_{\text{esc}}(R_\odot)^2}{u}\right)\\
        &\approx 1.1\cdot10^{33}\left(\frac{m\chi}{\text{MeV}}\right)^{-1}\text{s}^{-1}\,.
\end{align}
\end{subequations}
\subsection{Dark matter scattering rate inside the Sun}\label{sec: dmscatteringinthesun}
When the DM particle passes through the plasma of the Sun, it is possible to scatter with solar nuclei or electrons, assuming that there is a certain type of non-gravitational interaction between DM and SM particles. We calculate the total and differential scattering rates of DM with different targets in the Sun. We assume the scatterings are non-relativistic and follow the non-relativistic treatment in~\cite{An:2021qdl}. We first derive the general formulae for the scattering rates that are independent of the DM model. In the next section, we specifically consider fermionic DM with a dark-photon mediator.

We assume that the target particles follow a thermal Maxwell-Boltzmann velocity distribution at each local position characterized by the radial distance from the center of the Sun $r$,
\begin{equation}
    n_i(r)=n_i(r)\int\frac{\dd^3\vec{p}}{(2\pi)^3}f_i(p,r)\,,
\end{equation}
\begin{equation}\label{MBdistribution}
    f_i(p,r)=\left(\frac{2\pi}{m_i T(r)}\right)^{3/2}e^{-p^2/2m_i T(r)}\,.
\end{equation}
In the expression, the index $i$ represents different solar targets, i.e., electrons and various nuclear isotopes that make up the Sun; also, $n_i(r)$ is the number density of the $i$-th target at solar radius $r$, and $f_i(p,r)$ is the MB velocity distribution, which depends on the target's mass $m_i$, the target's momentum $p$, and the local temperature $T(r)$. The differential cross section in the non-relativistic limit is
\begin{equation}
    \dd\sigma=\frac{(2\pi)^4\delta^{(4)}(K_1+P_1-K_2-P_2)}{16m_i^2m_\chi^2v_{\text{rel}}}\frac{\dd^3\vec{k}_2}{(2\pi)^3}\frac{\dd^3\vec{p}_2}{(2\pi)^3}\langle\left|M_i\right|^2\rangle\,,
\end{equation}
where $K_1$ ($P_1$) and $K_2$ ($P_2$) are the four-momenta of the incoming and outgoing DM (target) particles. We always denote relativistic 4-vectors with capital letters. In the denominator, we apply the non-relativistic limit. The scattering rate $\Omega_i$ for a given target $i$ is
\begin{equation}\label{scatteringrate1}
\begin{split}
    \Omega_i=n_i \langle v_{\text{rel}} \sigma_{i}\rangle=\frac{n_i}{16m_i^2m_\chi^2}\int\frac{\dd^3\vec{p}_1}{(2\pi)^3}f_i(p_1)\int\frac{\dd^3\vec{k}_2}{(2\pi)^3}\int\frac{\dd^3\vec{p}_2}{(2\pi)^3}(2\pi)^4\delta^{(4)}(K_1+P_1-K_2-P_2)\langle\left|M_i\right|^2\rangle\,.
\end{split}
\end{equation}
We integrate out $\vec{p}_2$ using the spatial part of the delta function first, and write the energy conservation in terms of the classical kinetic energy,
\begin{equation}
\begin{split}
    \Omega_i=\frac{n_i}{16m_i^2m_\chi^2}\int\frac{\dd^3\vec{p}_1}{(2\pi)^3}f_i(p_1)\int\frac{\dd^3\vec{k}_2}{(2\pi)^3}2\pi\delta\left(\frac{k_1^2}{2m_\chi}+\frac{p_1^2}{2m_i}-\frac{k_2^2}{2m_\chi}-\frac{(\vec{k}_1+\vec{p}_1-\vec{k}_2)^2}{2m_i}\right)\langle\left|M_i\right|^2\rangle\,.
\end{split}
\end{equation}
Then we define the momentum transfer as $Q=K_2-K_1=(q^0,\vec{q})$ and change the variable $\vec{k}_2$ to $\vec{q}$,
\begin{equation}
\begin{split}
    \Omega_i=\frac{n_i}{16m_i^2m_\chi^2}\int\frac{\dd^3\vec{p}_1}{(2\pi)^3}f_i(p_1)\int\frac{\dd^3\vec{q}}{(2\pi)^3}2\pi\delta\left(q^2\left(\frac{1}{2m_\chi}+\frac{1}{2m_i}\right)+\frac{\vec{k}_1\cdot \vec{q}}{m_\chi}-\frac{\vec{p}_1\cdot \vec{q}}{m_i}\right)\langle\left|M_i\right|^2\rangle\,.
\end{split}
\end{equation}
In the non-relativistic limit and for the DM models we consider in this paper, the spin-averaged scattering matrix squared $\langle\left|M_i\right|^2\rangle$ only depends on $Q$. Then it is possible to analytically integrate over $\vec{p}_1$. It is easier to consider $\vec{p}_1$ in spherical coordinates with a $z$-axis oriented along the momentum transfer $\vec{q}$,
\begin{equation}
	\begin{split}
    &\Omega_i=\frac{n_i}{16m_i^2m_\chi^2}\int\frac{2\pi p_1^2\dd p_1\dd\cos\theta_{\vec{p}_1\vec{q}}}{(2\pi)^3}f_i(p_1)\int\frac{\dd^3\vec{q}}{(2\pi)^3}\\
	&\times2\pi\delta\left(q^2\left(\frac{1}{2m_\chi}+\frac{1}{2m_i}\right)+\frac{\vec{k}_1\cdot \vec{q}}{m_\chi}-\frac{p_1 q\cos\theta_{\vec{p}_1\vec{q}}}{m_i}\right)\langle\left|M_i(Q)\right|^2\rangle\,,
	\end{split}
\end{equation}
where $\theta_{\vec{p}_1\vec{q}}$ is the angle between $\vec{p}_1$ and $\vec{q}$. We integrate over $\cos\theta_{\vec{p}_1\vec{q}}$ and find
\begin{equation}\label{scatteringrate3}
    \Omega_i=\frac{n_i}{32\pi m_i^2m_\chi^2}\int_{p_{1,\mathrm{min}}}^{\infty}p_1^2\dd p_1\,f_i(p_1)\int\frac{\dd^3\vec{q}}{(2\pi)^3}\frac{m_i}{p_1 q}\langle\left|M_i(Q)\right|^2\rangle\,.
\end{equation}
Meanwhile, $p_1$ has a minimum value to ensure that the argument in the delta function can evaluate to zero in the range of $\cos\theta_{\vec{p}_1\vec{q}}\in[-1,1]$,
\begin{equation}\label{eq: p1min}
\begin{split}
    p_{1,\mathrm{min}}=\left|\frac{q}{2}\left(\frac{m_i}{m_\chi}+1\right)+\frac{m_i}{m_\chi}\frac{\vec{k}_1\cdot \vec{q}}{q}\right|=\left|\frac{q}{2}\left(\frac{m_i}{m_\chi}+1\right)+\frac{m_i}{m_\chi}k_1\cos\theta_{\vec{q}\vec{k}_1}\right|\,,
\end{split}
\end{equation}
where $\theta_{\vec{q}\vec{k}_1}$ is the angle between the spatial momentum transfer $\vec{q}$ and the initial momentum of the DM particle $\vec{k}_1$, while $p_{1,\mathrm{min}}$ physically is the minimum target momentum needed to make the momentum transfer $\vec{q}$ possible. Now we can substitute \cref{MBdistribution} and integrate $p_1$ in \cref{scatteringrate3},
\begin{equation}\label{eq: differential scattering rate}
\begin{split}
    \Omega_i=\frac{n_i}{32\pi m_i^2 m_\chi^2}\left(\frac{m_i}{2\pi T}\right)^{1/2}\int_{0}^{\infty}\dd q\int_{-1}^{1}\dd\cos\theta_{\vec{q}\vec{k}_1}\,q\exp\left(-\frac{p_{1,\mathrm{min}}^2}{2m_i T}\right)\langle\left|M_i(Q)\right|^2\rangle \,,
\end{split}
\end{equation}
where we write $\vec{q}$ in the spherical coordinate with $\vec{k}_1$ defining the $z$-axis. The momentum $p_{1,\mathrm{min}}$ depends on $\vec{q}$ and $\vec{k}_1$ as shown in \cref{eq: p1min}. Note that $\langle\left|M_i(Q)\right|^2\rangle$ may depend on the four-momentum $Q$. The temporal component $q^0$ is related to the spatial $\vec{q}$ by
\begin{equation}\label{eq: q0}
    q^0=\frac{\left|\vec{k}_1+\vec{q}\right|^2-k_1^2}{2m_\chi}=\frac{q^2+2k_1q\cos\theta_{\vec{q}\vec{k}_1}}{2m_\chi}\,.
\end{equation} 
\cref{eq: differential scattering rate} is the general expression for the non-relativistic scattering rate between the DM and the solar targets $i$.
We obtain the total scattering rate by summing over all target species, 
\begin{align}
    \Omega(r,v) & = \Omega_e + \sum_i \Omega_i\, , \label{eq: total scattering rate}
\end{align}
where the first term accounts for electrons and the term second for all nuclear isotopes present in the Sun.
In our simulations, we explicitly include the five most important nuclear isotopes as targets, \isotope{}{1}{H}, \isotope{}{4}{He}, \isotope{}{3}{He}, \isotope{}{16}{O}, and \isotope{}{56}{Fe}.

%%%%%%%%%%%%%%%%%%%%%%%%%%%%%%%%%%%%%%%%%%%%%%%%%%%%%%%%%%%%%%%
\subsection{Dark-photon-mediated dark matter scatterings in a medium}\label{sec: darkphotonmediator}
Now we consider the solar reflection of DM with a dark-photon mediator that is kinetically mixed with the hypercharge gauge boson. For small mediator masses ($\ll$keV), the scattering amplitude has a strong dependence on momentum transfer as it scales as $1/q^2$. Also, in a medium like the solar plasma, charged particles screen the electromagnetic interaction and attenuate the photon propagator. Since the dark photon is kinetically mixed with the SM~photon (at low energies), the screening effect on the SM photon will manifest itself as an attenuations of the mixing parameter, leading to an `effective' mixing parameter. The effect is significant when the momentum transfers $q$ are small ($\lesssim$~keV).

\subsubsection{Dark-photon model} 
We consider a dark photon $A'$ that is the vector boson of a dark $U(1)_D$ gauge group, which (at low energies) is kinetically mixed with the ordinary electromagnetic photon~\cite{Holdom:1985ag,Galison:1983pa}. The vacuum Lagrangian relevant for the low-energy phenomenology is given by
\begin{equation}
\begin{split}
    \mathcal{L}=-\frac{1}{4}F_{\mu\nu}F^{\mu\nu}-\frac{1}{4}F'_{\mu\nu}F'^{\mu\nu}-\frac{\epsilon}{2}F_{\mu\nu}F'^{\mu\nu}+\frac{1}{2}m_{A'}^2A'_\mu A'^\mu+eA_\mu J^\mu_{\text{EM}}+g_D A'_\mu J^\mu_D\,.
\end{split}
\end{equation}
The mixing term allows the dark photon to couple to the electromagnetic current of the SM with the coupling strength $\epsilon e$. This can be seen by changing to the basis that diagonalizes the kinetic term. In particular, the DM-proton and the DM-electron interactions are generated through exchange of the dark photon. The relevant terms in the Lagrangian are
\begin{equation}
    \mathcal{L}=\epsilon e A_\mu'(\bar{p}\gamma^\mu p-\bar{e}\gamma^\mu e)\,.
\end{equation}

For DM-electron interactions, the matrix element squared is~\cite{Essig:2011nj,Essig:2015cda}
\begin{equation}\label{meq}
    \langle\left|M_e(q)\right|^2\rangle=\frac{16\epsilon^2e^2g_D^2m_e^2m_\chi^2}{(q^2+m_{A'}^2)^2}=\langle\left|M_e(q_\text{ref})\right|^2\rangle\left|F_{\text{DM}}(q)\right|^2\,,
\end{equation}
where $q_{\text{ref}}$ is the reference momentum transfer, often taken to be $q_{\text{ref}}=\alpha m_e$, and the DM form factor $F_{\text{DM}}(q)$ is defined as
\begin{equation}
    F_{\text{DM}}(q)=\frac{q_{\text{ref}}^2+m_{A'}^2}{q^2+m_{A'}^2}\simeq\left\{
    \begin{aligned}
    &1\,,&\quad m_{A'}\gg q_\text{ref}\\
    &\frac{q_{\text{ref}}^2}{q^2}\,,&\quad m_{A'}\ll q_\text{ref}
    \end{aligned}
    \right.\,.
\end{equation}
The reference DM-electron cross section is defined as
\begin{equation}\label{refcse}
    \overline\sigma_e\equiv\frac{16\pi\alpha\alpha_D\epsilon^2\mu_{\chi e}^2}{(q_{\text{ref}}^2+m_{A'}^2)^2}\simeq\left\{
    \begin{aligned}
    \frac{16\pi\alpha\alpha_D\epsilon^2\mu_{\chi e}^2}{m_{A'}^4}\,,\quad m_{A'}\gg q_\text{ref}\\
    \frac{16\pi\alpha\alpha_D\epsilon^2\mu_{\chi e}^2}{q_{\text{ref}}^4}\,,\quad m_{A'}\ll q_\text{ref}
    \end{aligned}
    \right.\,,
\end{equation}
where $\alpha$ is the fine structure constant, $\mu_{\chi e}$ is the DM-electron reduced mass and $\alpha_D\equiv g_D^2/4\pi$. The relation between the reference cross section and the matrix element squared is
\begin{equation}
    \overline\sigma_e\equiv\frac{\mu_{\chi e}^2\langle\left|M_e(q_{\text{ref}})\right|^2\rangle}{16\pi m_\chi^2 m_e^2}\,.
\end{equation}
The reference DM-proton cross section is defined as
\begin{equation}
    \overline\sigma_p\equiv\frac{16\pi\alpha\alpha_D\epsilon^2\mu_{\chi p}^2}{(q_{\text{ref}}^2+m_{A'}^2)^2}\,,
\end{equation}
and it has a simple relation with the reference DM-electron cross section,
\begin{equation}\label{eq: link_DM_p_DM_e}
    \frac{\overline\sigma_p}{\overline\sigma_e}=\left(\frac{\mu_{\chi p}}{\mu_{\chi e}}\right)^2\,.
\end{equation}

We cannot use the naive expression \cref{meq} in \cref{eq: differential scattering rate} to obtain the differential and total scattering rates, but instead we first need to include charge screening within the solar medium.

\subsubsection{In-medium effects}\label{subsubsec: in-medium effects}

\begin{figure}[t]
    \centering
    \includegraphics[width=0.45\textwidth]{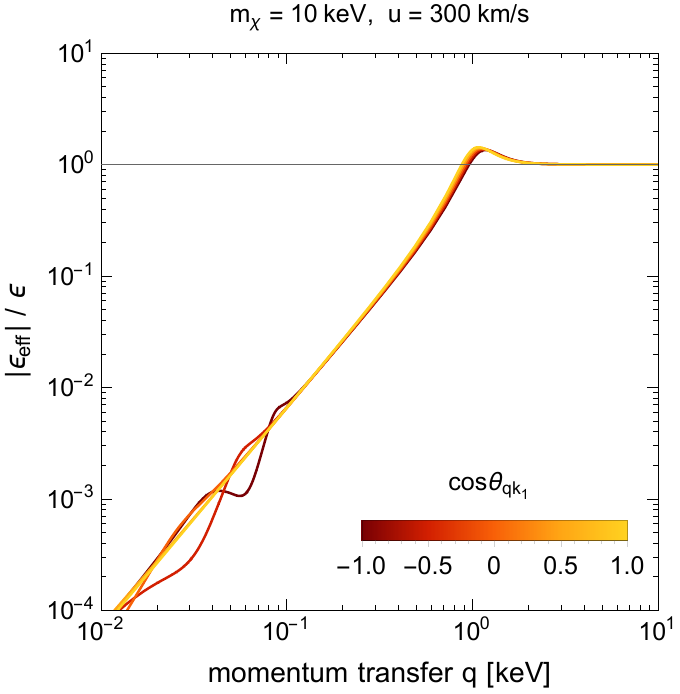}
    \caption{The ratio between $\left|\epsilon_{\text{eff}}\right|$ and bare coupling $\epsilon$ from \cref{eq: in medium effect basic}, evaluated at $r=0.5R_\odot$. We use a DM mass of 10~keV and a velocity $w(r)=\sqrt{u^2+v_{\text{esc}}(r)^2}$ where the asymptotic speed is $u=300\;\text{km s}^{-1}$.}
    \label{fig: effective_coupling}
\end{figure}

In the medium, the photon polarization tensor modifies the kinetic mixing $\epsilon$ in the vacuum to an effective kinetic mixing~\cite{Hochberg:2015fth}
\begin{equation}\label{eq: in medium effect basic}
    \epsilon_{\text{eff}}=\epsilon\frac{Q^2}{Q^2-\Pi_{T,L}(Q)}\simeq\epsilon\frac{q^2}{q^2+\Pi_{T,L}(Q)}\,,
\end{equation}
where $Q=(q^0,\vec{q})$ is the four-momentum transfer of a scattering process, and $\Pi_{T,L}$ is the self-energy for a transverse and longitudinal mode photon respectively, defined as
\begin{equation}\label{poltensor}
    \Pi^{\mu\nu}=\Pi_T\sum_{i=1,2}\epsilon_i^{T\mu}\epsilon_i^{T*\nu}+\Pi_L\epsilon^{L\mu}\epsilon^{L\nu}\,,
\end{equation}
where $\epsilon^{T,L}$ are polarization vectors for transverse and longitudinal modes,
\begin{align}
    \epsilon^L&=\frac{1}{\sqrt{Q^2}}(q,q^0\frac{\vec{q}}{q})\,,\\
    \epsilon^T_{1,2}&=\frac{1}{\sqrt{2}}(0,1,\pm i,0)\,.
\end{align}
For non-relativistic scatterings, the scattering through the longitudinal mode dominates~\cite{Hochberg:2015fth}. In this limit, we have
\begin{equation}\label{scatteringmatrix}
\begin{split}    \langle\left|M_e(q)\right|^2\rangle\simeq\frac{16\epsilon^2e^2g_D^2m_e^2m_\chi^2}{(q^2+m_{A'}^2)^2}\left|\frac{q^2}{q^2+\Pi_L(Q)}\right|^2=\langle\left|M_e(q_\text{ref})\right|^2\rangle\left|F_{\text{DM}}(q)\right|^2\frac{q^4}{\left|q^2+\Pi_L(Q)\right|^2}\,,
\end{split}
\end{equation}
for DM-electron scatterings.
For the longitudinal mode self-energy $\Pi_L(q)$, we use the result in~\cite{DeRocco:2022rze}. For a review of the detailed calculation, we refer to \cref{app: inmedium}. The electrons in the Sun follow a Maxwell-Boltzmann velocity distribution with standard deviation $\sigma=\sqrt{T/m_e}$. The plasma frequency is given by $\omega_p^2=e^2n_e/m_e$. The longitudinal mode self-energy is given by
\begin{equation}\label{longitudinalinmediumpol}
    \Pi_L(Q)\simeq\omega_p^2\frac{m_e}{q^0}\xi(Z(\xi-\delta)-Z(\xi+\delta))\,,
\end{equation}
where $\xi(Q)=\frac{q^0}{\sqrt{2}\sigma q}$ and $\delta(Q)=\frac{-Q^2}{2\sqrt{2}qm_e\sigma}$. The function $Z(\xi)$ is defined as
\begin{equation}
    Z(\xi)=\frac{1}{\sqrt{\pi}}\int_{-\infty}^\infty \dd x\frac{e^{-x^2}}{x-\xi}=\sqrt{\pi}e^{-\xi^2}(i-Erfi(\xi))\,.
\end{equation}
This is the electron's contribution to the longitudinal photon self-energy. For the other charged components' contributions to the self energy, the calculation is identical, as explained in \cref{app: inmedium}. We include both electronic and nuclear contribution to the photon self-energy. \cref{fig: effective_coupling} shows an example of the ratio between $\abs{\epsilon_{\text{eff}}}$ and the bare coupling $\epsilon$ based on \cref{eq: in medium effect basic} as a function of the momentum transfer $q$, evaluated at $r=0.5R_\odot$. We set the DM mass to be 10~keV and assume an initial velocity of  $w(r)=\sqrt{u^2+v_{\text{esc}}(r)^2}$, where $u$ is the asymptotic speed of the DM particle far away from the Sun, taken to be a typical value of $u=300\;\text{km s}^{-1}$ (specifying these  parameters is necessary to fix $q^0$ according to \cref{eq: q0}). We can see the effective coupling is heavily screened for $q\lesssim1$ keV, then for $q \sim 1$~keV it is slightly enhanced ($\epsilon_{\text{eff}}/\epsilon > 1$) since $\text{Re}\left[\Pi_L\right]$ is negative, while for $q\gtrsim2$~keV the in-medium effects are negligible.

\subsubsection{Dark matter-electron scatterings}
Now we have the matrix element squared \cref{scatteringmatrix} of DM-electron interactions, which takes in-medium effect into account. Applying it to the general equation of the scattering rate \cref{eq: differential scattering rate}, we have the expression for the DM-electron scattering rate $\Omega_e$ in the Sun as an integral over $q$ and $\theta_{\vec{q}\vec{k}_1}$,
\begin{equation}\label{scatteringrate5}
\begin{split}
    \Omega_e=\frac{n_e e^2 g_D^2 \epsilon^2}{2\pi}\left(\frac{m_e}{2\pi T}\right)^{1/2}\int_{0}^{\infty}\dd q\int_{-1}^{1}\dd\cos\theta_{\vec{q}\vec{k}_1}\,q\exp\left(-\frac{p_{1,\mathrm{min}}^2}{2m_e T}\right)\frac{1}{(q^2+m_{A'}^2)^2}\frac{q^4}{\left|q^2+\Pi_L(q)\right|^2}\,,
\end{split}
\end{equation}
where
\begin{equation}
    p_{1,\mathrm{min}}=\left|\frac{q}{2}\left(\frac{m_e}{m_\chi}+1\right)+\frac{m_e}{m_\chi}k_1\cos\theta_{\vec{q}\vec{k}_1}\right|\,.
\end{equation}
We can also express the scattering rate \cref{scatteringrate5} in terms of the scattering cross section $\overline\sigma_e$ as
\begin{equation}\label{scatteringrate2}
\begin{split}
    \Omega_e=\frac{\overline\sigma_e n_e}{2\mu_{\chi e}^2}\left(\frac{m_e}{2\pi T}\right)^{1/2}\int_{0}^{\infty}\dd q\int_{-1}^{1}\dd\cos\theta_{\vec{q}\vec{k}_1}\,q\exp\left(-\frac{p_{1,\mathrm{min}}^2}{2m_e T}\right)\left|F_{\text{DM}}(q)\right|^2\frac{q^4}{\left|q^2+\Pi_L(q)\right|^2}\,,
\end{split}
\end{equation}
which is more convenient when one wants to evaluate the scattering rates with $\overline\sigma_e$ as an input value. 

\subsubsection{Dark matter-nuclei scatterings} Through the dark-photon mediator, the DM-nucleus cross section $\sigma_N$ is related to the DM-proton cross section $\sigma_p$ as follows,
\begin{equation}\label{DM-NandDM-pcs}
    \frac{\dd\sigma_N}{\dd q^2}=\frac{\dd\sigma_p}{\dd q^2}Z^2\left|F_N(q)\right|^2\,,
\end{equation}
where $Z$ is the total number of charge of the nucleus, and $F_N(q)$ is the nuclear form factor. One can verify that this treatment is equivalent to considering the Coulomb scattering with the composite particle carrying charge of $Ze$ with an extra form factor $F_N(q)$. Thus, we can obtain the DM-nucleus scattering rate $\Omega_N$ simply by generalizing \cref{scatteringrate5},
\begin{equation}
\begin{split}
    &\Omega_N=\frac{n_N Z^2e^2 g_D^2 \epsilon^2}{2\pi}\left(\frac{m_N}{2\pi T}\right)^{1/2}\int_{0}^{\infty}\dd q\int_{-1}^{1}\dd\cos\theta_{\vec{q}\vec{k}_1}\\
    &\times q\exp\left(-\frac{p_{1,\mathrm{min}}^2}{2m_N T}\right)\frac{1}{(q^2+m_{A'}^2)^2}\frac{q^4}{\left|q^2+\Pi_L(q)\right|^2}\left|F_N(q)\right|^2\,,
\end{split}
\end{equation}
where $n_N$ is the local number density of nucleus $N$, $m_N$ is the nuclear mass, and
\begin{equation}
    p_{1,\mathrm{min}}=\left|\frac{q}{2}\left(\frac{m_N}{m_\chi}+1\right)+\frac{m_N}{m_\chi}k_1\cos\theta_{\vec{q}\vec{k}_1}\right|\,.
\end{equation}
Meanwhile, for low-mass DM particles, the momentum transfer is small enough that the mediator probes the nucleus as a whole and $F_N(q)\simeq 1$. We connect the scattering rate $\Omega_N$ to the observable $\overline\sigma_p$,
\begin{equation}\label{scatteringrateN}
\begin{split}
    \Omega_N=\frac{\overline\sigma_p Z^2 n_N}{2\mu_{\chi p}^2}\left(\frac{m_N}{2\pi T}\right)^{1/2}\int_{0}^{\infty}\dd q\int_{-1}^{1}\dd\cos\theta_{\vec{q}\vec{k}_1}\,q\exp\left(-\frac{p_{1,\mathrm{min}}^2}{2m_N T}\right)\left|F_{\text{DM}}(q)\right|^2\frac{q^4}{\left|q^2+\Pi_L(q)\right|^2}\,.
\end{split}
\end{equation}
In the dark photon model, the observable $\overline\sigma_p$ is linked to $\overline\sigma_e$ through \cref{eq: link_DM_p_DM_e}.

\begin{figure}[t]
    \centering
    \includegraphics[width=\textwidth]{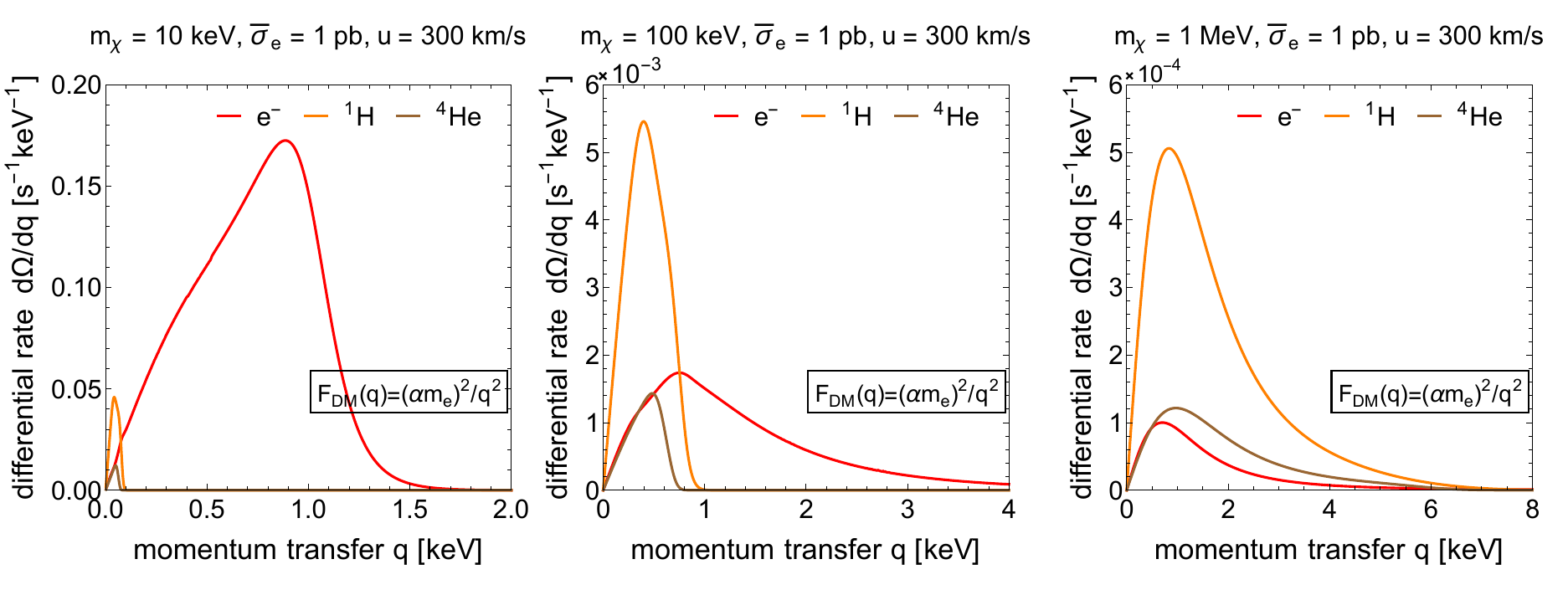}
    \caption{The differential scattering rate $\dd\Omega/\dd q$ for different targets in the Sun (electrons, protons, and $^4$He) versus the momentum transfer $q$, after integrating out the angle $\cos\theta_{\vec{q}\vec{k}_1}$, evaluated at $r=0.5R_\odot$, with DM initial speed $w(r)=\sqrt{u^2+v_{\text{esc}}(r)^2}$, and $u=300\;\text{km s}^{-1}$. The DM scatters via an ultralight dark photon and the reference DM-electron cross section is set to $\overline\sigma_e=1$~pb.}
    \label{fig: dsigmadqatr0.5}
\end{figure}

\begin{figure}[t]
    \centering
    \includegraphics[width=\textwidth]{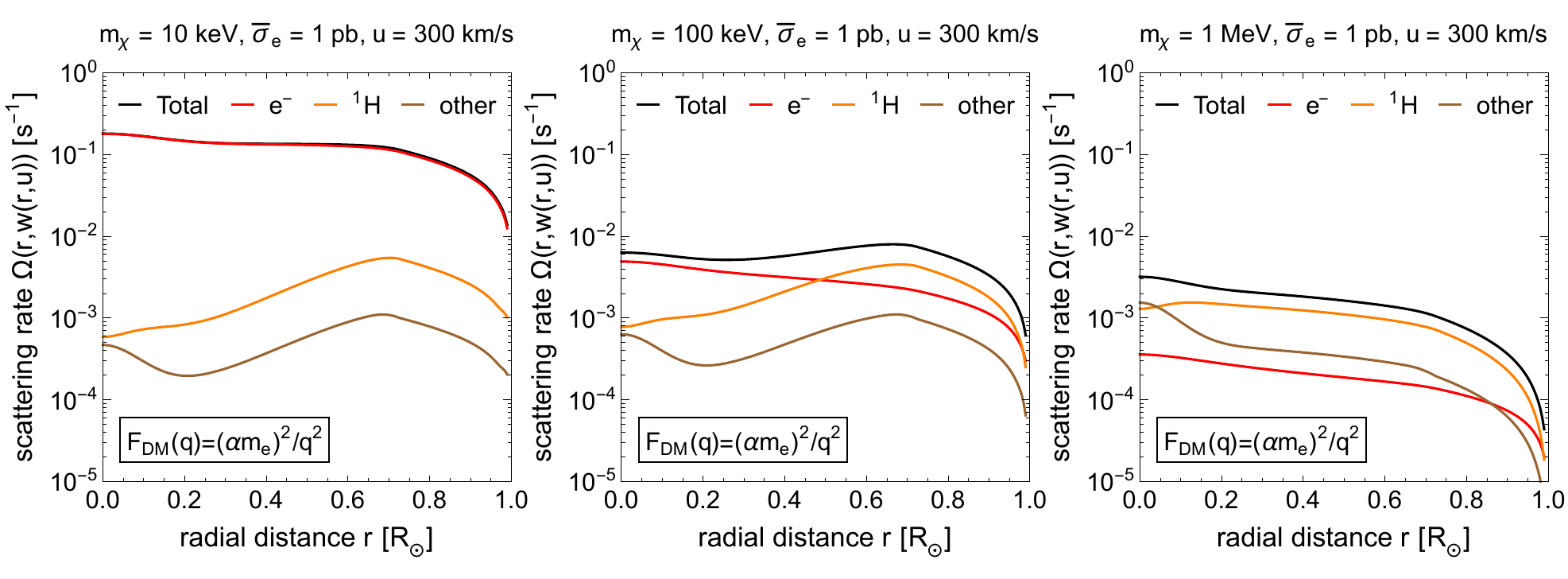}
    \caption{Total scattering rate $\Omega$ for different targets in the Sun (electrons, protons, and all other nuclei combined) versus the radial distance from the center of the Sun $r$, with DM initial speed $w(r)=\sqrt{u^2+v_{\text{esc}}(r)^2}$, and $u=300\;\text{km s}^{-1}$. DM scatters through an ultralight dark photon and the reference DM-electron cross section is set to $\overline\sigma_e=1$~pb. Contributions from nuclei other than protons are shown in brown demonstrating that proton scatterings dominate the contribution from nuclei.}
    \label{fig: total rate plot}
\end{figure}

In \cref{fig: dsigmadqatr0.5}, we show the differential scattering rate $\dd\Omega/\dd q$ in the case of an ultralight dark-photon mediator, after integrating over the angle $\cos\theta_{\vec{q}\vec{k}_1}$ in \cref{scatteringrate2} and \cref{scatteringrateN}, evaluated at a radius $r=0.5R_\odot$. We can see that for lower DM masses, such as 10~keV, the scatterings with an electron can result in keV-scale momentum transfers, leading to a change in the DM's velocity of $\sim0.1c$. For larger DM masses (closer to $\sim$1~MeV), the scattering of DM with nuclei becomes more important. However, heavier DM gets less acceleration, for example, a momentum, transfer of $q\sim$1~keV for a 1~MeV DM particle results in a change of velocity of only $\sim10^{-3}c$. Moreover, the DM-electron scattering rate is also lower for larger DM masses (for a fixed reference cross section), because the rate $\Omega\propto e^2 g_D^2 \epsilon^2 \propto \overline\sigma_e/\mu_{\chi e}^2$. For these two reasons, the sensitivity to the SRDM flux degrades as the DM mass is increased from the keV to MeV scale. 

We also show the scattering rate of DM with different targets as a function of radius $r$ in \cref{fig: total rate plot} with the same choice of $w(r)$ as in \cref{fig: dsigmadqatr0.5}. We  see that the DM scattering off electrons dominates over the scattering off nuclei for lower DM masses, while the DM scattering off protons dominates at higher DM mass. Moreover, regardless of the DM mass, the proton is the most important nuclear target for most values of the radius $r$.

\begin{figure}[t]
    \centering
    \includegraphics[width=0.45\textwidth]{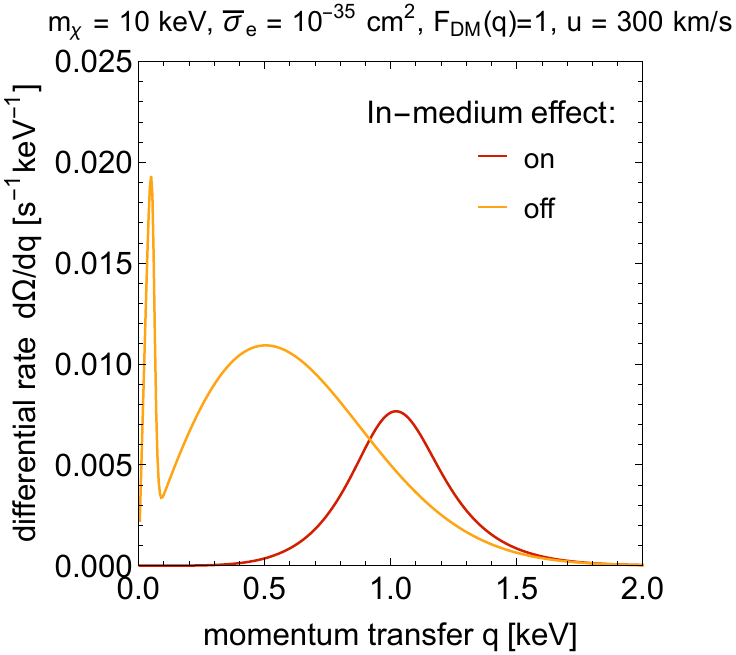}
    \caption{The comparison of the differential scattering rate $\dd\Omega/\dd q$ for DM interacting with a heavy dark-photon mediator ($F_{\rm DM}=1$) with and without the inclusion of in-medium effects in the Sun, after integrating over the angle $\cos\theta_{\vec{q}\vec{k}_1}$ for the parameter choices given in the plot.}
    \label{fig: heavy mediator in-medium effect}
\end{figure}

Although the focus of this work is on DM with an ultralight mediator, we revisit the calculation of DM with a heavy dark-photon mediator by taking into account the in-medium effect, as  previous studies of solar reflection of DM with a heavy dark-photon mediator neglected these effects~\cite{An:2017ojc, Emken:2017hnp, Emken:2021lgc, An:2021qdl}. \cref{fig: heavy mediator in-medium effect} shows the comparison of the differential scattering rate with and without in-medium effects in the Sun for a DM with a heavy dark photon mediator. When ignoring the in-medium effects, we see a substantial contribution from DM-nucleus scattering, corresponding to the spikes at very low $q$. When we include in-medium effects, the DM-nucleus scattering is totally screened at low $q$. The DM-electron scattering rate is reduced and its peak is shifted to higher $q$ as well. We will present the impact of the in-medium effects on the reflection flux in \cref{sec: results-flux-heavy}. 

\subsection{Direct detection of solar-reflected dark matter}\label{sec: dd of srdm}
We focus on DM direct-detection experiments with noble liquid or semiconductor targets, which are sensitive to DM-electron scattering.  
We first discuss, however, DM-nucleus scattering to set some notation, and since it is an instructive example.

\subsubsection{Nuclear recoil spectrum}
Given some generic differential DM particle flux $\frac{\dd\Phi_\chi}{\dd v_\chi}$ passing through a detector with $N_T$ nuclei, the resulting nuclear recoil spectrum is given by
\begin{equation}
    \frac{dR}{dE_R}=N_T\int_{v_\chi>v_{\text{min}}}\dd v_\chi\frac{\dd\Phi_\chi}{\dd v_\chi}\frac{\dd\sigma_{N}}{dE_R}\,,
\end{equation}
where $v_{\text{min}}$ is the minimum DM velocity to produce a nuclear recoil with energy $E_R$,
\begin{equation}
    v_{\text{min}}=\sqrt{\frac{m_N E_R}{2\mu_{\chi N}^2}}\,.
\end{equation}
In this work, the DM flux consists of the halo DM flux $\Phi_\text{halo}$ and the SRDM flux $\Phi_\odot$, hence
\begin{equation}\label{nuclearrecoilspec}
    \frac{dR}{dE_R}=N_T\int_{v_\chi>v_{\text{min}}}\dd v_\chi\left(\frac{\dd\Phi_\text{halo}}{\dd v_\chi}+\frac{\dd\Phi_\odot}{\dd v_\chi}\right)\frac{\dd\sigma_{N}}{dE_R}\,.
\end{equation}
Our focus in this paper is on DM whose mass is so small that even the fastest DM particles in the halo do produce detectable recoils in direct-detection experiments. In this case, the first term in \cref{nuclearrecoilspec} can be ignored and only the SRDM contributes to the event rate. Below, we thus do not distinguish $\Phi_\chi$ and $\Phi_\odot$ and use the same notation $\Phi_\chi$ for the flux from SRDM.
For DM with a dark-photon mediator, the DM-nucleus cross section can be written from \cref{DM-NandDM-pcs} as 
\begin{equation}
    \frac{\dd\sigma_N}{\dd E_R}=\frac{m_p\overline\sigma_p}{2\mu_{\chi p}^2v_\chi^2}Z^2\abs{F_{\text{DM}}(q)}^2\abs{F_N(q)}^2\,,
\end{equation}
where $q=\sqrt{2m_NE_R}$. Using \cref{nuclearrecoilspec}, we have the nuclear recoil spectrum,
\begin{equation}\label{nuclearrecoilspec2}
\begin{split}
    \frac{\dd R}{\dd E_R}=N_T\frac{m_p\overline\sigma_p}{2\mu_{\chi p}^2}Z^2\abs{F_{\text{DM}}(q)}^2\abs{F_N(q)}^2\int_{v_\chi>v_{\text{min}}}\dd v_\chi\frac{1}{v_\chi^2}\frac{\dd\Phi_\chi}{\dd v_\chi}\,.
\end{split}
\end{equation}

\subsubsection{Atomic ionization}\label{sec: atomic excitation}
DM particles can ionize atoms through DM-electron scattering~\cite{Essig:2011nj}. The differential scattering rate off an electron with principal quantum number $n$ and orbital angular momentum $l$ in an atom in a target with $N_T$ atoms is~\cite{Essig:2011nj,Essig:2012yx,Essig:2015cda}
\begin{equation}
\begin{split}
    \frac{\dd R_\text{ion}^{nl}}{\dd\ln E_e}=N_T\frac{\overline\sigma_e}{8\mu_{\chi e}^2}\int q\dd q\,\abs{F_{\text{DM}}(q)}^2\abs{f_{\text{ion}}^{nl}(k',q)}^2\int_{v_\chi>v_{\text{min}}}\dd v_\chi\frac{1}{v_\chi^2}\frac{\dd\Phi_\chi}{\dd v_\chi}\,,
\end{split}
\end{equation}
where $E_e=k'^2/2m_e$ is the energy for free final-state electron, and $f_\text{ion}^{nl}$ is the ionization form factor for the $(n,l)$-orbital. The scattering is inelastic, since the initial-state electron is bounded with binding energy $E_B^{nl}$. The minimum velocity is
\begin{equation}
    v_{\text{min}}=\frac{\Delta E}{q}+\frac{q}{2m_\chi}\,,
\end{equation}
where $\Delta E=\abs{E_B^{nl}}+E_e$ is the energy transfer from the DM particle to the electron.

For the calculations of the energy spectra, projected sensitivities, and exclusion limits from xenon targets, we use the xenon form factors $f_{\text{ion}}^{nl}(k',q)$ from~\cite{Essig:2012yx}. However, we will investigate the impact of using other form factors in \cref{app: xenon form factors}, in particular, using 
\texttt{DarkARC}~\cite{Catena:2019gfa} and \texttt{AtMolDM}~\cite{Hamaide:2021hlp}. 

\subsubsection{Crystal excitations}\label{sec: crystal excitation}
For DM-electron scattering in semiconductors (crystals), the DM can excite an electron from the valence to the conduction band. Our focus will be on silicon, although our results apply more generally to other crystals. The DM-electron differential scattering rate with a total energy deposit of $\Delta E=E_e$ in a crystal of $N_{\text{cell}}$ unit cells is given by~\cite{Essig:2015cda}
\begin{equation}
    \begin{split}
        \frac{\dd R_{\text{crys}}}{\dd\ln E_e}=N_{\text{cell}}\alpha\frac{\overline\sigma_e m_e^2}{\mu_{\chi e}^2}\int\frac{\dd q}{q^2}E_e\abs{F_{\text{DM}}(q)}^2\abs{f_{\text{crys}}(E_e,q)}^2\int_{v_\chi>v_{\text{min}}}\dd v_\chi\frac{1}{v_\chi^2}\frac{\dd\Phi_\chi}{\dd v_\chi}\,,
    \end{split}
\end{equation}
where $f_\text{crys}(E_e,q)$ is the crystal form factor, and
\begin{equation}
    v_{\text{min}}=\frac{E_e}{q}+\frac{q}{2m_\chi}\,.
\end{equation}
Given an energy deposit of $E_e$, the probability of exciting $Q$ electron-hole pairs (often simply referred to as ``electrons'') in the crystal target is written as $p(Q, E_e)$. The rate for obtaining charge yield $Q$ is then 
\begin{equation}
    \Delta R_Q=\int\dd E_e \frac{\dd R}{\dd E_e}p(Q, E_e)\,.
\end{equation}
We use the ionization yield model from~\cite{Ramanathan:2020fwm} at 100~K.

For the calculations of the energy spectrum, projected sensitivities, and exclusion limits from a silicon target, we use the silicon form factors for boosted DM-electron scattering from~\cite{Essig:2024ebk}. This uses the silicon form factor from \texttt{QCDark}~\cite{Dreyer:2023ovn,QCDark}, which is a DFT-based calculation and performs ab-initio all-electron calculations, i.e.,  it treats core and valence electrons on the same footing, rather than approximating the effects of core electrons with pseudopotentials. Moreover, it takes into account the collective effects in silicon targets by multiplying the crystal form factor squared $\left|f_\text{crys}(E_e,q)\right|^2$ by $1/\left|\epsilon(E_e,q)\right|^2$,
\begin{equation}
    \left|f_\text{crys}(E_e,q)\right|^2 \rightarrow \frac{\left|f_\text{crys}(E_e,q)\right|^2}{\left|\epsilon(E_e,q)\right|^2},
\end{equation}
where $\epsilon(E_e,q)$ is the dielectric function~\cite{Knapen:2021run, Hochberg:2021pkt}. This captures the plasmon peak, which can be accessed in the scatters of boosted DM.  For the dielectric function, we use the Lindhard model~\cite{lindhard1954properties}. In \cref{app: silicon form factors}, we will also compare the energy spectra from SRDM scatters to those when using $\left|f_\text{crys}(E_e,q)\right|^2$  by itself without correcting it with the dielectric function, as well as when using the form factor from \texttt{QEDark}~\cite{Essig:2015cda,QEDark}, both with and without the inclusion of the dielectric function.
%%%%%%%%%%%%%%%%%%%%%%%%%%%%%%%%%%%%%%%%%%%%%%%%%%%%%%%%%%%%%%%
\section{Monte Carlo simulations of the solar-reflected DM flux}\label{sec: montecarlosimulations}
We calculate predictions of the SRDM~flux through Earth using MC~simulations of DM~particle trajectories through the solar medium.
In this study, we use and extend the MC~tool Dark Matter Simulation Code for Underground Scatterings -- Sun Edition (\texttt{DaMaSCUS-SUN}~\cite{Emken2021}), which was released alongside~\cite{Emken:2021lgc}.
We expand the capabilities of~\texttt{DaMaSCUS-SUN} to include in-medium effects and ultralight mediators or long-range interactions.
In this section, we summarize  the simulation of a single DM~particle as it passes through the Sun, scatters inside the hot plasma, and escapes the star with increased kinetic energy.
For a more detailed description, we will refer to specific parts of~\cite{Emken:2021lgc} and \cref{app: simulation details}.

The first step is to sample the initial conditions, i.e., the initial position and velocity of a DM~particle from the galactic halo that is going to intersect the solar surface.
The initial velocity~$\mathbf{u}$ does not simply follow the Maxwell-Boltzmann distribution of the SHM, previously introduced in \cref{eq: SHM velocity distribution}, because of the Sun's motion in the galactic rest frame.
We also have to account for the two facts that (a) faster particles enter the Sun at a greater rate, and (b) gravitational focusing can pull in slow particles from a greater volume.
The initial velocity distribution is derived in Appendix~A4 of~\cite{Emken:2021lgc} and given by
\begin{align}
f_{\mathrm{IC}}(\mathbf{u}) &=\mathcal{N}_{\mathrm{IC}}\left(u+\frac{v_{\mathrm{esc}}\left(R_{\odot}\right)^2}{u}\right) f_{\chi}(\mathbf{u})\, , \label{eq: initial velocity distribution}
\end{align}
where $\mathcal{N}_{\mathrm{IC}}$ is some normalization constant.
The initial position~$\mathbf{x}_0$ has to be chosen very far away from the Sun, where its gravitational potential has not yet affected the kinematic distribution of the~DM.
In practice, we sample the initial position at a distance of about 1000~AU from the Sun.
In order to ensure that the particle will hit the Sun, we can derive the maximum impact parameter,
\begin{align}
    b_\mathrm{max} &= \sqrt{1+\frac{v_\mathrm{esc}(R_\odot)^2}{u}} R_\odot\, ,
\end{align}
where the fact that $b_\mathrm{max} > R_\odot$ encapsulates the gravitational focusing.
On its path towards the Sun, the particle follows a hyperbolic Kepler orbit, as described in Appendix~A2 of~\cite{Emken:2021lgc}.
We use the analytic solution to the Kepler problem to propagate the particle to the direct vicinity of the solar surface, right before the particle enters the solar medium.

Once the DM~particle enters the Sun, its orbit is no longer Keplerian, and we have to solve the equations of motion, summarized in Appendix~A1 of~\cite{Emken:2021lgc}, numerically.
We use the Runge-Kutta-Fehlberg method, an adaptive, iterative algorithm for the numerical solution of ordinary differential equations~\cite{Fehlberg1969}.\footnote{A brief review of the RKF algorithm can be found in Appendix~A3 of~\cite{Emken:2021lgc}.}
At each time step of the numerical solution, we check if the particle has scattered.
The probability of scattering within a time interval~$\Delta t$ along a trajectory is given by
\begin{align}
P(\Delta t)=1-\exp \left(-\int_0^{\Delta t} \frac{\dd t}{\tau(r(t), v(t))}\right)\, ,
\end{align}
where we defined the mean free time~$\tau(r, v)$ in terms of the total scattering rate given by \cref{eq: total scattering rate},
\begin{align}
    \tau(r, v)=\Omega(r, v)^{-1}\, .
\end{align}
At every time step of the RKF method, we need to evaluate the mean free time and therefore the total scattering rate.
This can involve the evaluation of complicated functions and numerical integration of ill-behaved integrands, which can become the time bottleneck of the simulation.
To speed up the simulations, we tabulate the total scattering rate as a function of radial distance~$r$ and DM~speed~$v_\chi$ and use bilinear interpolation during the trajectory simulation.
To locate the next scattering event, we sample a uniformly distributed random number~$\xi\in(0,1)$ and solve~$P(\Delta t)=\xi$ along the DM~particle's path.

Once we identified the location of a scattering, we identify the target. The probability to scatter on target species~$i$ is given by 
\begin{align}
    P(\text{scattering on target } i)=\frac{\Omega_i\left(r, v_\chi\right)}{\Omega\left(r, v_\chi\right)}\, .
\end{align}
So far, these steps have been identical to the description in~\cite{Emken:2021lgc}.
The next step is to find the new DM~velocity after the scattering, where we change the random variables to be sampled.
Instead of sampling the target velocity~$\mathbf{v}_T$ and scattering angle~$\alpha$ in the center-of-mass frame (as we did in~\cite{Emken:2021lgc}), we sample the momentum transfer~$q$ and angle~$\cos \theta$ between the initial DM~velocity $\mathbf{v}_\chi$ and the momentum transfer $\mathbf{q}$.
This choice of variable is particular beneficial in the case of light mediators.

The joint probability density function~(PDF) of $(q,\cos\theta)$ for a target~$i$ is given by the differential scattering rate of \cref{eq: differential scattering rate},
\begin{align}
    f_i(q,\cos\theta) &= \frac{1}{\Omega_i(r,v)} \frac{\dd^2\Omega_i}{\dd q \dd\cos\theta}\, , \label{eq: PDF q cos theta}
\end{align}
where~$\Omega_i(r,v)$ is the total scattering rate and ensures the normalization of the PDF. The region for sampling is $\cos\theta\in\left[-1,1\right]$ trivially, while the momentum transfer $q$ has an upper limit $q_{\text{max}}$,
\begin{equation}\label{eq: qmax}
    q_{\text{max}}=2\mu_{\chi i}\left(v_\chi+N\sqrt{\frac{2T(r)}{m_i}}\right)\,,
\end{equation}
where $\mu_{\chi i}$ is the reduced mass of the DM particle and the target, $v_\chi$ is the initial DM speed, $T(r)$ is the temperature in the Sun at radius $r$, and $N$ is a parameter we set manually. To reach $q=q_\text{max}$, the target's initial speed needs to be no less than $N\sqrt{2T(r)/m_i}$, whose probability is suppressed by $\exp(-N^2)$ according to the Maxwell-Boltzmann velocity distribution. Hence we can safely neglect momentum transfers larger than $q_\text{max}$. We choose $N=5$ in our practical simulation. 

Drawing random samples from the distribution in \cref{eq: PDF q cos theta} can be inefficient in particular for light mediators, where only a small part of the domain carries most probability. Furthermore, the evaluation of the PDF requires the calculation of the total scattering rate as the normalization constant, which can be computationally expensive.
We circumvent this problem by sampling using the Metropolis algorithm.
The Metropolis algorithm is a Markov Chain Monte Carlo (MCMC) method to draw random samples from a distribution that does not require the distribution to be normalized.
A detailed description of our method to sample~$(q,\cos\theta)$ can be found in \cref{app: simulation details}.

In combination with a uniformly distributed azimuthal angle~$\phi\in(0,2\pi)$, we can construct the full momentum transfer vector~$\mathbf{q}$, and thereby also the new DM~velocity after the scattering via
\begin{align}
    \mathbf{v}_\chi \rightarrow \mathbf{v}_\chi^\prime \equiv \mathbf{v}_\chi + \frac{\mathbf{q}}{m_\chi}\, .
\end{align}
Having identified the DM~particle's velocity after the scattering, we can continue the trajectory simulation and repeat the previous steps.

The simulation of a DM~trajectory continues until one of the following exit conditions applies:
\begin{enumerate}
    \item The particle leaves the solar plasma with~$v_\chi > v_\mathrm{esc}(R_\odot)$. If this particle has scattered at least once, it is regarded as \textit{reflected}. If the particle has not scattered, we consider it a \textit{free} particle as its kinetic energy has not changed.
    \item We consider a DM~particle as \textit{captured}, if it either scatters too many times (in practice we abort the simulation of a trajectory after $10^4$ scatterings), or it propagates along a bound orbit without scattering for a long time (we abort after $10^8$ time steps of the RKF method). These arbitrary choices of upper limits slightly overestimate the number of captured particles.
\end{enumerate}
During data generation, we count all free, reflected, and captured particles.
In addition, we propagate the reflected DM~particles to a distance of 1~AU using the analytic solution to the Kepler problem, where we record the particle's speed.
We repeat simulating particles until we generated a sufficiently large set of speed data points that will allow us to estimate the differential SRDM~particle flux,
\begin{subequations}
\label{eq: differential SRDM flux MC}
\begin{align}
    \frac{\dd \Phi_{\odot}}{\dd v_\chi}&=\frac{1}{4 \pi \ell^2} \frac{\dd \mathcal{R}_{\odot}}{\dd v_\chi}\, ,
    \intertext{where we used the differential reflection spectrum}
    \frac{\dd \mathcal{R}_{\odot}}{\dd v_\chi} &\equiv \mathcal{R}_{\odot} f_{\odot}^{\mathrm{KDE}}\left(v_\chi\right)\, ,
    \intertext{and the total reflection rate}
    \mathcal{R}_{\odot} &\equiv \frac{N_{\mathrm{refl}}}{N_{\mathrm{sim}}} \Gamma\left(m_\chi\right) \, .
\end{align}
\end{subequations}
Here, $N_\mathrm{sim}$ and $N_\mathrm{refl}$ are the total number of simulated and reflected particles respectively,~$\ell = 1\text{ AU}$ is the distance between the Earth and the Sun, and~$\Gamma(m_\chi)$ is the rate of halo~DM particles falling into the Sun, given by \cref{eq: infalling rate}.
The speed distribution of the reflected flux is given by~$f_\odot^\mathrm{KDE}(v_\chi)$, a Kernel Density Estimation (KDE) based on the recorded speed data.
Kernel Density Estimation is a non-parametric procedure to estimate probability densities based on data~\cite{rosenblatt1956,parzen1962}.\footnote{A brief review of KDE and bias mitigation at domain boundaries can be found in Appendix~A of~\cite{Emken:2018run}.}
Compared to simple histograms, KDEs have the benefit of yielding smooth distributions, and in our case smooth SRDM~spectra.

\begin{figure}[t]
    \centering
    \includegraphics{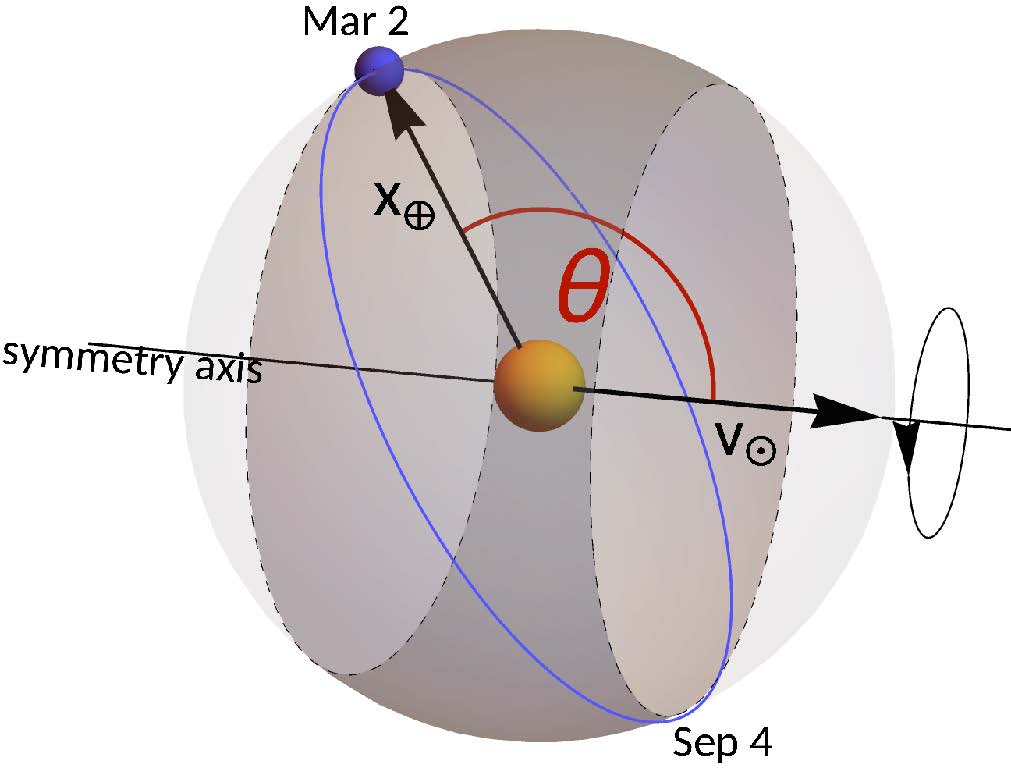}
    \caption{The symmetry axis along the Sun's velocity and the isoreflection angle $\theta$, defined in \cref{eq: isoreflectionangle}. Along the Earth's orbit, it approximately covers the interval $\theta\in\left[60^\circ,120^\circ\right]$, reaching the maximum and minimum on Mar.~2 and Sep.~4 respectively. Figure taken from~\cite{Emken:2021lgc}}.
    \label{fig: isoreflectionangle}
\end{figure}
Our code is also capable of simulating the SRDM flux from different directions to determine the anisotropy of the SRDM flux and the annual modulation of the signal rate. The velocity distribution of halo DM in the galactic rest frame is isotropic. As the Sun is moving with velocity $\mathbf{v}_\odot$ in the galactic rest frame, the DM velocity distribution in the solar rest frame is boosted along the opposite of $\mathbf{v}_\odot$, resulting in the ``DM wind.'' The spherical symmetry of the original distribution in the galactic rest frame breaks down to an axisymmetry with a symmetry axis along $\mathbf{v}_\odot$, as illustrated in \cref{fig: isoreflectionangle}. We want to determine whether the anisotropy of the in-coming DM in the solar frame gets ``washed away" after solar reflection, or if the SRDM flux preserves a certain amount of anisotropy. We introduce the isoreflection angle $\theta$ in \cref{fig: isoreflectionangle}, defined as the angle between the Sun's velocity $\mathbf{v}_\odot$ and the Earth's location $\mathbf{x}_\oplus$ related to the Sun,
\begin{equation}\label{eq: isoreflectionangle}
    \theta\equiv\sphericalangle(\mathbf{x}_\oplus, \mathbf{v}_\odot)\,,
\end{equation}
which is useful to quantify the anisotropies of SRDM. For directions with the same isoreflection angle~$\theta$, the SRDM flux is the same because of the axisymmetrical distribution of the incoming DM. The Earth covers the isoreflection angle between $\sim60^\circ$ (around September 4) and $\sim120^\circ$ (around March 2) along its orbit. The possible dependence of the SRDM flux on $\theta$ leads to a new type of annual modulation of potential DM signals in direct detection. We present the modulation of the SRDM flux in \cref{sec: anisoflux} and the signal rate in \cref{sec: modsignalrate}.
%%%%%%%%%%%%%%%%%%%%%%%%%%%%%%%%%%%%%%%%%%%%%%%%%%%%%%%%%%%%%%%
\section{Results}
\label{sec: results}
In this section, we present the main results of our work. First, we show the simulation result of the SRDM flux for DM with an ultralight dark-photon mediator, investigate its general properties, and consider the limits for strong and weak interaction cross sections. With the simulation results for the SRDM flux in hand, we calculate the event rates induced by the SRDM flux in different detectors and calculate the resulting exclusion limits from current direct-detection experiments.  We also present projections for planned and proposed detectors. Moreover, we update previous results for DM that interacts through a heavy dark-photon mediator by now including the in-medium effects in the Sun.

\subsection{SRDM flux spectrum for dark matter coupled to an ultralight dark photon}\label{sec: results-flux}
\subsubsection{Solar reflection spectra}
\begin{figure}[t]
    \centering
    \includegraphics[width=0.63\textwidth]{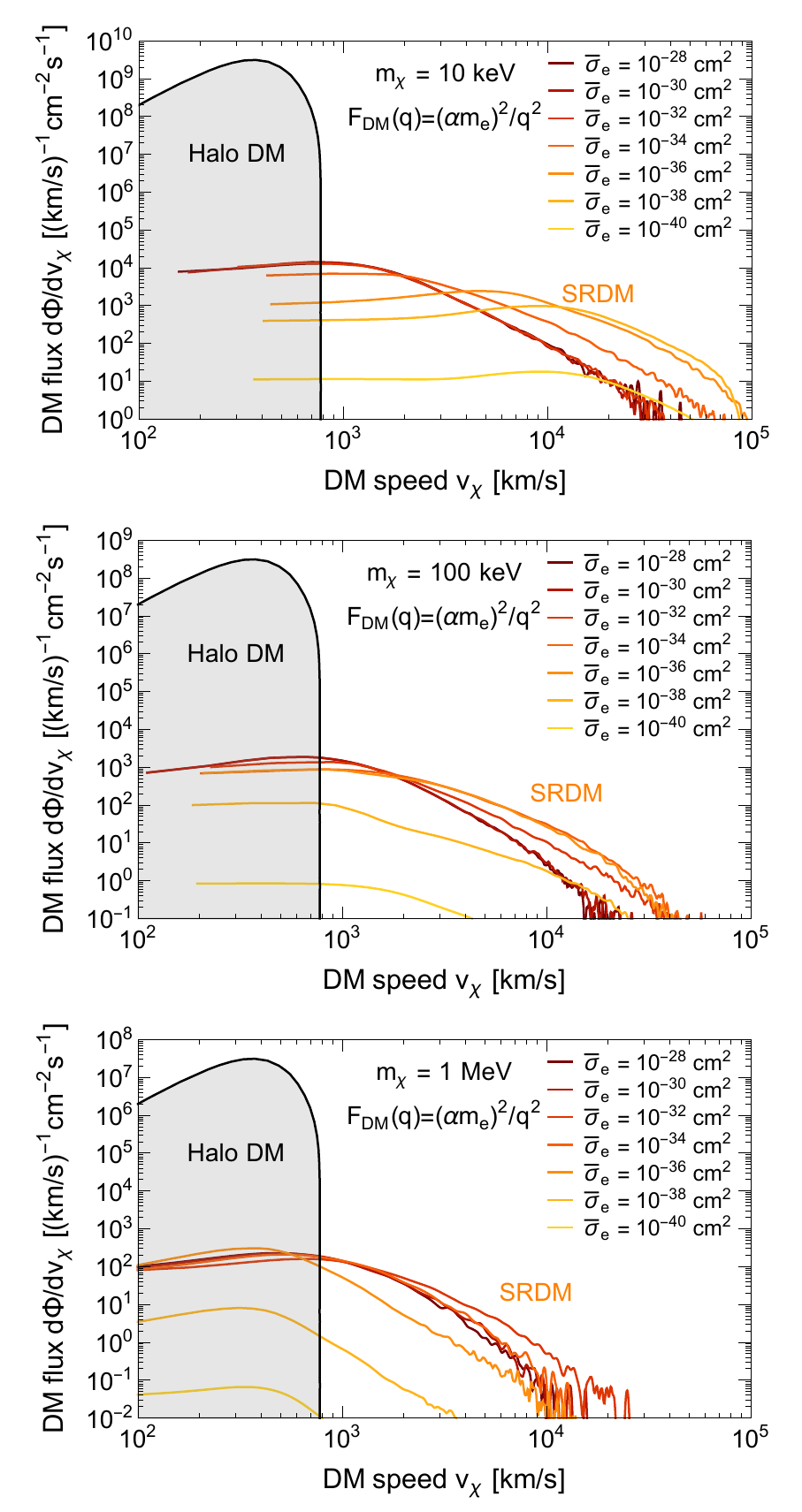}
    \caption{The differential SRDM flux for DM interacting with an ultralight dark-photon mediator for $m_\chi=10$~keV (\textbf{top}), $m_\chi=100$~keV (\textbf{middle}), and $m_\chi=1$~MeV (\textbf{bottom}) and for a wide range of DM-electron scattering cross sections $\overline\sigma_e$  (changing from dark red to yellow for decreasing cross section values).  The flux from halo DM is shown with a black curve.}
    \label{fig: threefluxplot}
\end{figure}
We use the MC simulation to estimate the differential flux $\frac{\dd\Phi_\chi}{\dd v_\chi}$ of SRDM, where $v_\chi$ is the DM speed, at a distance of 1 AU from the Sun. The SRDM flux for DM with an ultralight dark-photon mediator and the comparison with the flux from halo DM is shown in \cref{fig: threefluxplot} for three DM masses, namely $m_\chi=10$~keV (top plot), $m_\chi=100$~keV (middle plot), and $m_\chi=1$~MeV (bottom plot) and for a wide range of DM-electron scattering cross sections $\overline\sigma_e$. We can see that the total SRDM flux is orders of magnitude lower than the halo DM flux. However, the velocity of the DM particles in the SRDM flux extend far beyond the halo DM's escape velocity. 

We can see from \cref{fig: threefluxplot} that DM particles with larger masses are less boosted than DM particles with lower masses, which agrees with intuition and our evaluation of the differential scattering rates in \cref{fig: dsigmadqatr0.5}. For larger cross sections $\overline\sigma_e$ (dark red curves for $\overline\sigma_e\gtrsim10^{-30}\;\text{cm}^2$), the Sun is essentially opaque to the DM particles, and almost all scatterings happen at the outer and cooler layers, resulting in very similar reflected spectra for even larger cross sections. For lower cross sections, the DM particles can reach closer to the center of the Sun, where the  temperature is higher, resulting in fluxes that contain more energetic DM particles. For very low cross sections, the Sun becomes transparent to  scatterings, and the reflected flux is attenuated. In this case, most reflected particles only scatter once in the Sun, with the scattering probability directly proportional to the cross section. Hence, for very low cross sections, the normalization of the reflected flux scales approximately with the cross section.

\subsubsection{Gravitational capture of dark matter}
\begin{figure}[t]
    \centering
    \includegraphics[width=0.6\textwidth]{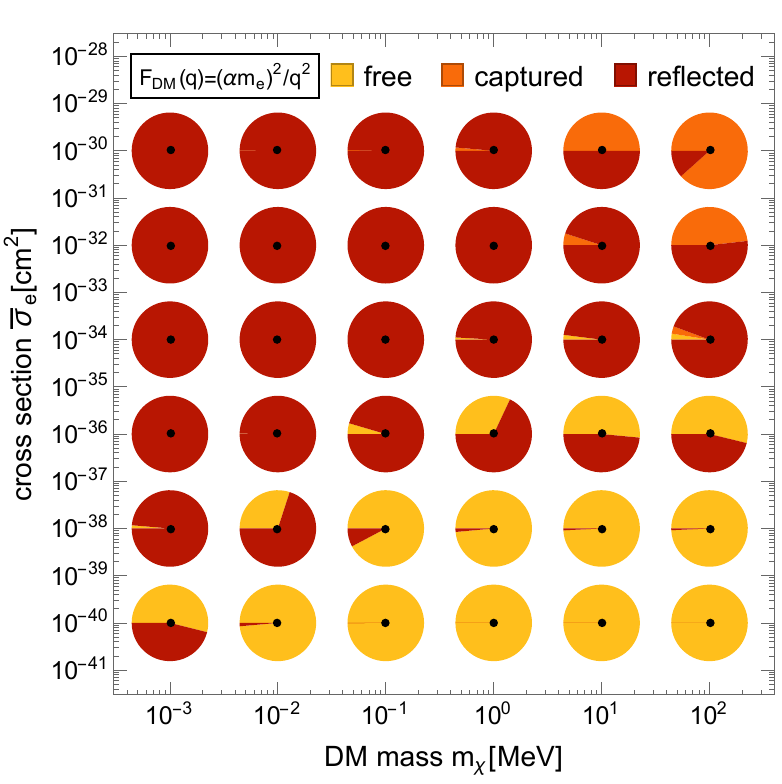}
    \caption{Pie-charts for the relative proportions of the DM particles entering the Sun and that pass through without scattering (``free'', yellow), get gravitationally captured (orange), or get reflected (red), as a function of DM mass $m_\chi$ and the interaction cross section $\overline\sigma_e$.}
    \label{fig: gravcap}
\end{figure}
As discussed in \cref{sec: montecarlosimulations}, our simulation code is capable of describing the gravitational capture of DM particles by the Sun. In particular, it is possible for the DM particles to be down-scattered in energy and become gravitationally bounded and ``captured'' by the Sun, especially for larger DM masses and larger cross sections, where the scatterings are less likely to accelerate the DM particles sufficiently to escape the Sun's gravitational well.

 In \cref{fig: gravcap}, we show the relative proportions of free, captured, and reflected particles, as a function of DM mass and interaction cross section, for DM coupled to an ultralight dark photon. We can see that the gravitational capture is significant for DM mass $m_\chi\gtrsim10\;\text{MeV}$, accompanied by sufficiently large interaction cross sections.
\subsubsection{Anisotropy of solar reflection}\label{sec: anisoflux}
\begin{figure}[t]
\centering
  \includegraphics[width=0.47\linewidth]{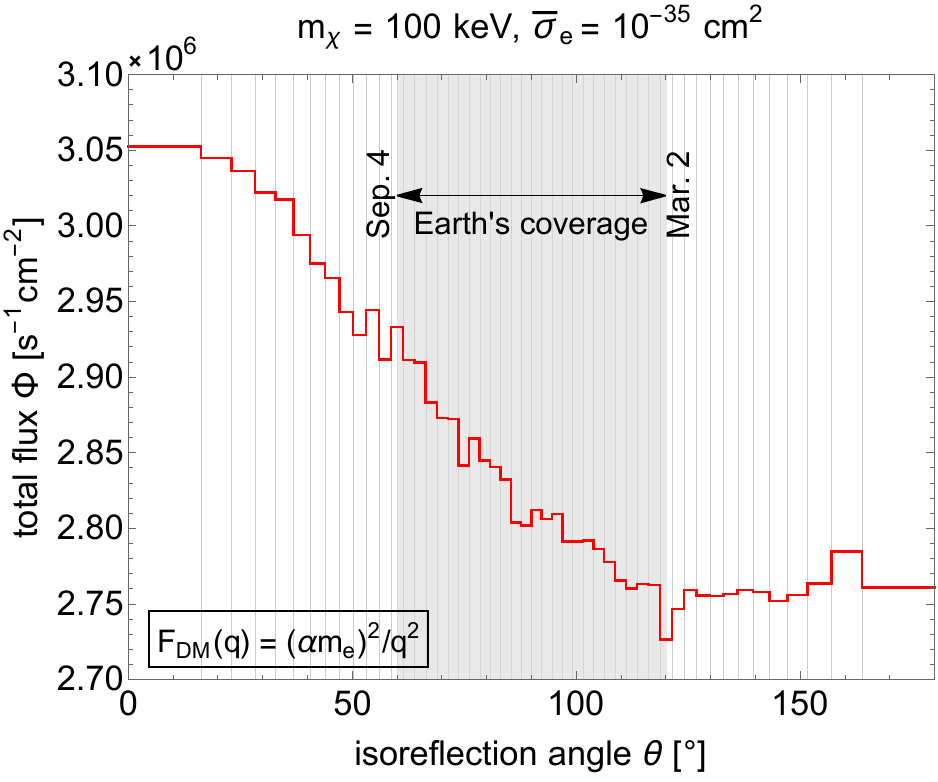}
  \includegraphics[width=0.47\linewidth]{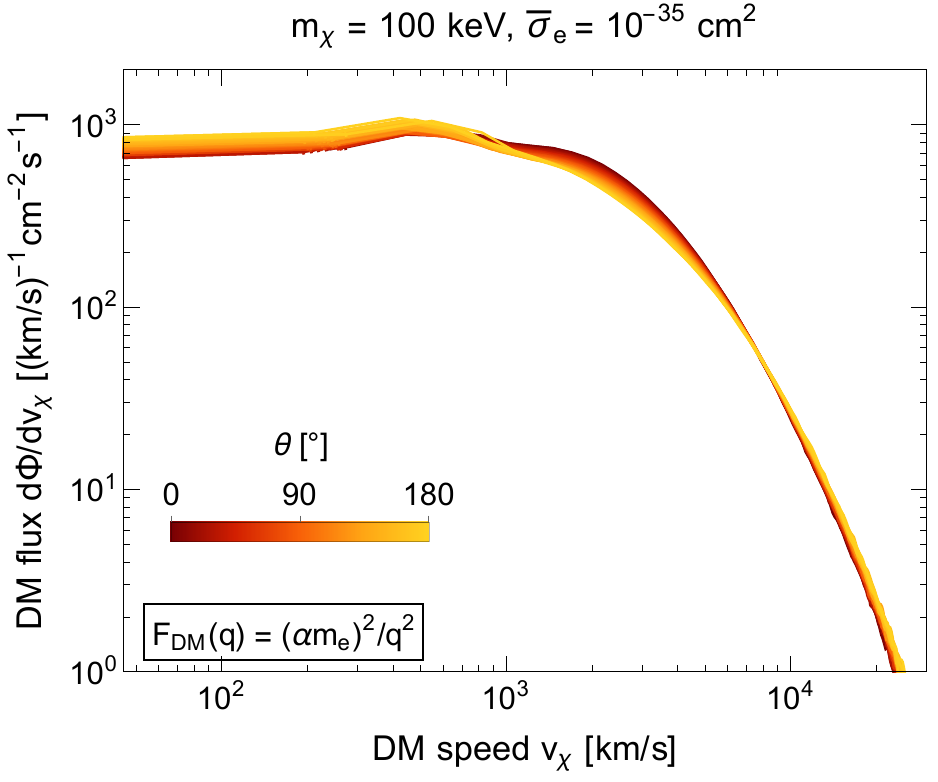}
\caption{Anisotropy of the SRDM flux. The left plot shows the SRDM flux $\Phi$ as a function of isoreflection angle $\theta$, defined as the angle between the Sun's velocity $\mathbf{v}_\odot$ and the Earth's location $\mathbf{x}_\oplus$ relative to the Sun, see \cref{eq: isoreflectionangle}. The gray area is the interval of $\theta$ that the Earth covers through its orbit in the solar system. The right plot shows the differential SRDM flux $\dd\Phi/\dd v_\chi$ for different isoreflection angles.}
\label{fig: anisoflux_and_diffflux}
\end{figure}

In \cref{sec: montecarlosimulations}, we discussed the anisotropy of the SRDM flux and the isoreflection angle $\theta$ (in particular, see \cref{fig: isoreflectionangle}). We compute the SRDM flux using 50~isoreflection rings. At each isoreflection ring, the flux is evaluated at a fixed distance of 1~AU from the Sun (we do not consider for now the fluctuation of the radius of the Earth's orbits around the Sun, but see below).  The directional dependence of the SRDM flux $\Phi$ on the isoreflection angle $\theta$ is shown in \cref{fig: anisoflux_and_diffflux} (left), for a DM mass of 100~keV and an interaction cross section of $\overline\sigma_e=10^{-35}\;\text{cm}^2$. The area shaded in gray is the interval of isoreflection angle that the Earth covers over the course of one year. We find that the flux can deviate from the mean value by up to 3\% if observed from the Earth, or around 5\% overall in the solar system. We also show the differential SRDM flux along different isoreflection angles in \cref{fig: anisoflux_and_diffflux} (right). The high-energy tail of the fluxes have a weaker dependence on the isoreflection angle than the total flux. Since it is the high-energy tail that contributes to the event rate in direct-detection experiments with eV-scale threshold for sub-MeV DM particles, we do not expect a strong modulation of the signal rate in this case. We show the modulation of the SRDM signal rate for DM coupled to an ultralight dark photon for a silicon target in \cref{sec: modsignalrate}.

\begin{table}
\begin{tabular}{|c|c|c|}
\hline In-medium effect&on&off\\
\hline $\langle$nScatterings$\rangle$&1.02&1.83\\
\hline $\langle r\rangle$(last scattering) [$r_\odot$]&0.513&0.568\\
\hline $\langle r\rangle$(deepest scattering) [$r_\odot$]&0.495&0.545\\
\hline free particles [\%]&40.5&32.1\\
\hline reflected particles [\%]&59.4&67.8\\
\hline captured particles [\%]&0.1&0.1\\
\hline
\end{tabular}
\caption{Comparison of simulated results with and without the inclusion of dark-photon in-medium effects in the Sun, for a DM mass $m_\chi = 10$~keV, and a cross section $\overline\sigma_e=10^{-35}\;\text{cm}^2$, for a heavy dark-photon mediator. We list the average  number of scatterings for all simulated particles, the average radius of where the last scattering in the Sun occurs, and the average of where the deepest scattering occurs.}
\label{table: heavy IME flux compare}
\end{table}
\begin{figure}[t]
    \centering
    \includegraphics[width=0.45\textwidth]{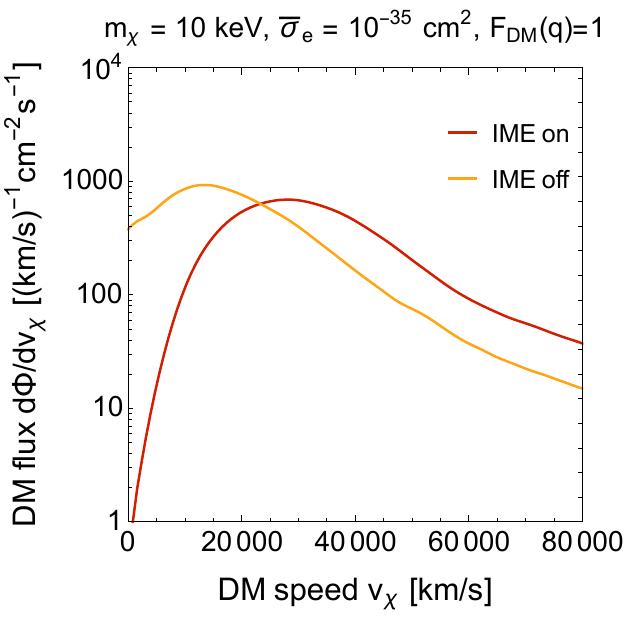}
    \caption{Comparison of the SRDM flux with (red) and without (orange) the inclusion of dark-photon in-medium effects in the Sun.}
    \label{fig: heavy IME flux compare}
\end{figure}

\subsection{SRDM flux spectrum for dark matter coupled to a heavy dark photon}\label{sec: results-flux-heavy}
We revisit the results for DM with a heavy dark-photon mediator first presented in~\cite{Emken:2021lgc}. We now include in-medium effects for DM scattering in the Sun also for a heavy mediator, which changes the scattering rates and hence impacts the exclusion limits and projections. We compare the simulation results with and without in-medium effects in \cref{table: heavy IME flux compare} for $m_\chi = 10$~keV and $\overline{\sigma}_e=10^{-35}$~cm$^2$. Despite there being fewer reflected particles for the case that in-medium effects are included, the DM particles make it deeper into the Sun, where the temperature is larger, because of the absence of low-$q$ DM-nucleus scatterings (see \cref{fig: heavy mediator in-medium effect}). We compare the SRDM fluxes in \cref{fig: heavy IME flux compare}. The flux with in-medium effects included has more particles with a higher velocity. As a result, by considering the in-medium effects in the Sun, we expect stronger constraints from direct-detection experiments than previous results that neglected in-medium effects~\cite{An:2017ojc, Emken:2017hnp, Emken:2021lgc, An:2021qdl}.

\subsection{Direct-detection results for dark matter coupled to an ultralight dark-photon}
In this section, we present direct-detection limits from the SRDM component from existing experiments, as well as projections for upcoming and proposed detectors, for DM coupled to an ultralight dark-photon mediator. We also present the annual modulation of a potential SRDM signal.
For our calculations, we perform a parameter scan in the $(m_\chi,\overline\sigma_e)$-plane and calculate the $p$-value at each point. For a given confidence level (CL), the boundary for each excluded region is defined by $p=1-\mathrm{CL}$.
%%%%%%%%%%%%%%%%%%%%%%%%%%%%
\subsubsection{Exclusion limits}\label{sec: lightmediatorbound}
\begin{figure}[t]
    \centering
    \includegraphics[width=0.6\textwidth]{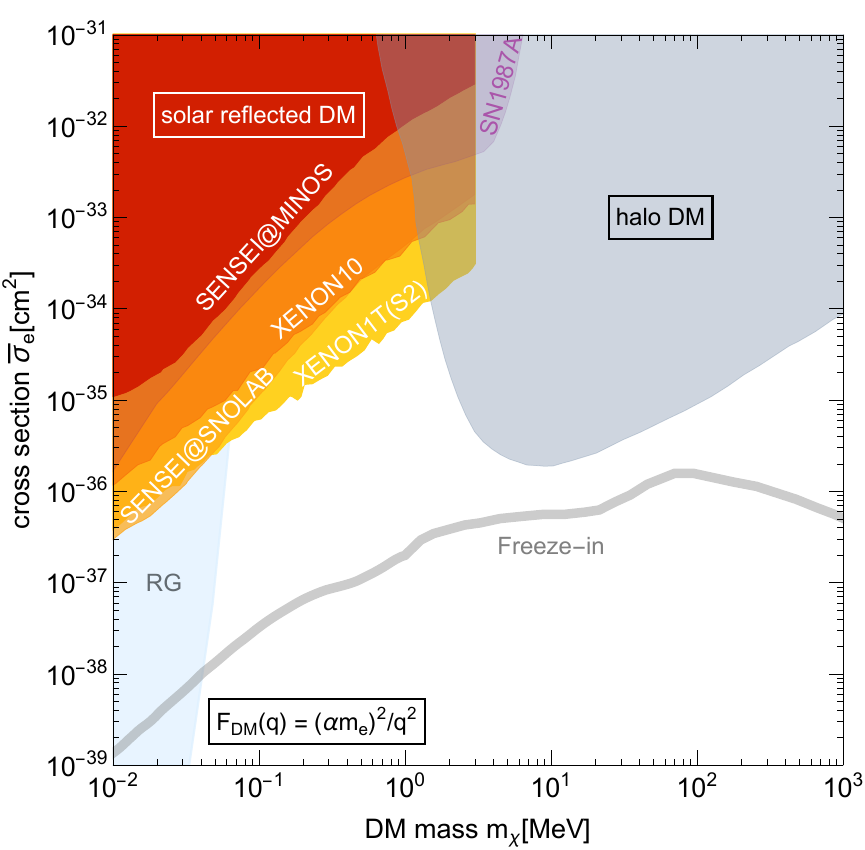}
    \caption{Solar reflection exclusion limits (90\% CL) from direct-detection experiments for DM coupled to an ultralight dark photon. The excluded regions shaded in solar colors are from SENSEI@MINOS~\cite{SENSEI:2020dpa}, SENSEI@SNOLAB~\cite{SENSEI:2023zdf}, XENON10~\cite{Angle:2011th, Essig:2012yx, Essig:2017kqs}, and XENON1T(S2-only)~\cite{XENON:2019gfn}. The shaded purple region is the constraint from SN1987A~\cite{Chang:2018rso}, and the region shaded in light blue on the left is the red giants cooling constraint~\cite{Vogel:2013raa}. For halo DM, the excluded region is the combined limit from SENSEI@SNOLAB~\cite{SENSEI:2023zdf} and SENSEI@MINOS~\cite{SENSEI:2020dpa}, while within the contour there are other limits from, e.g., DAMIC-M~\cite{Aguilar-Arevalo:2019wdi} and XENON10~\cite{Essig:2017kqs}. The freeze-in benchmark is plotted in gray~\cite{Essig:2011nj,Chu:2011be,Essig:2015cda,Dvorkin:2019zdi}.}
    \label{fig: lightmediatorbound}
\end{figure}
We show in \cref{fig: lightmediatorbound} with ``solar colors'' (shades of red, orange, and yellow) the 90\% CL limits on the SRDM flux component using data from SENSEI@MINOS~\cite{SENSEI:2020dpa}, SENSEI@SNOLAB~\cite{SENSEI:2023zdf}, XENON10~\cite{Angle:2011th, Essig:2012yx, Essig:2017kqs} and XENON1T (S2-only)~\cite{XENON:2019gfn}.\footnote{The solar-reflection limits from the SENSEI@SNOLAB data were already shown in~\cite{SENSEI:2023zdf} based on the SRDM simulation results discussed in this paper.}  The current exclusion limits from the halo DM component are shown with the shaded gray region in \cref{fig: lightmediatorbound}, which is a combination of the SENSEI@SNOLAB~\cite{SENSEI:2023zdf} and SENSEI@MINOS~\cite{SENSEI:2020dpa} limits. The stellar constraints from SN1987A~\cite{Chang:2018rso} and red giants cooling~\cite{Vogel:2013raa} are shown with the shaded light purple and light blue regions, respectively. Along the ``freeze-in benchmark''~\cite{Essig:2011nj,Chu:2011be,Essig:2015cda,Dvorkin:2019zdi} that is plotted in gray, the DM relic abundance is obtained from the freeze-in mechanism~\cite{Hall:2009bx}. We can see that the SRDM component extends the reach of direct-detection experiments to lower DM masses than traditional searches for halo DM. Vast regions in the parameter space for sub-MeV DM are excluded by considering SRDM in existing direct-detection experiments, surpassing even the bound from SN1987A~\cite{Chang:2018rso}.

\begin{figure}[t]
    \centering
    \includegraphics[width=\textwidth]{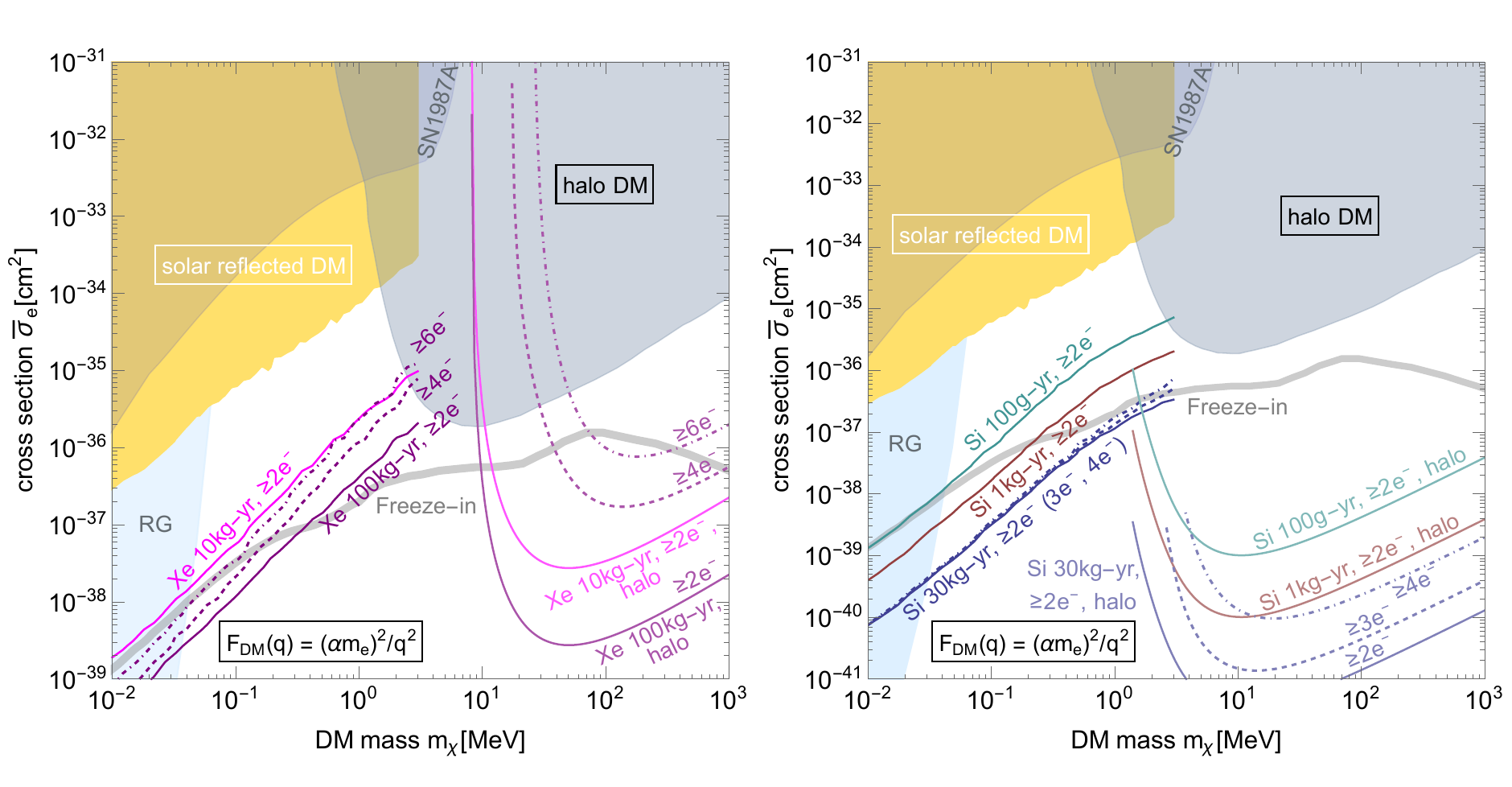}
    \caption{Projections for the solar-reflected DM flux for upcoming and proposed DM detectors with xenon (left) or silicon (right) targets, for DM coupled to an ultralight dark photon. The left plot shows SRDM projections for xenon detectors for a 10~kg-yr exposure and a threshold of 2~electrons (magenta curve), 100~kg-yr and a threshold of 2~electrons, 4~electrons, or 6~electrons (solid, dashed, dot-dashed purple curves). The right plot shows SRDM projections for silicon detectors with a 100~g-yr exposure and a threshold of 2 electrons (green curve), 1 kg-yr and 2 electrons (red curve), 30 kg-yr and 2, 3, or 4 electrons (solid, dashed, dot-dashed blue curves). The yellow region is the combined SRDM limits from \cref{fig: lightmediatorbound} and the other lines and shaded regions are as in \cref{fig: lightmediatorbound}.}
    \label{fig: lightmediatorprojections}
\end{figure}

For xenon, we use the calculations discussed in \cref{sec: atomic excitation}, while for silicon we use the calculations discussed in \cref{sec: crystal excitation}.  We compare the constraints when using different xenon form factors in \cref{app: xenon form factors} and when using different silicon crystal form factors in \cref{app: silicon form factors}.  
Note that for XENON10 (S2-only), we use the same electron yield and S2 signal modeling as in~\cite{Essig:2017kqs} as well as the same S2 binning. For XENON1T (S2-only), we use the same analysis\footnote{https://github.com/XENON1T/s2only\_data\_release} as in~\cite{XENON:2019gfn} to calculate the S2 signals from the energy spectrum. 
The S2 bins for XENON1T (S2-only) are 50 photoelectrons (PE) wide,  from 150 to 3000~PE. We find that the S2 bins higher than 500~PE have minimal impact on constraining SRDM (at least for light mediators). We have also tested that different binning settings do not have significant impact on the constraint. 
For SENSEI@MINOS, we show the best limit obtained from considering each of the $Q=1$, $Q=2$, ..., $Q=6$ electron-bins, while for SENSEI@SNOLAB we use a single bin that combines events with $Q=4-10$~electrons. (The official SENSEI limits shown in~\cite{SENSEI:2023zdf} use a likelihood ratio test, although the final limits are very similar.)

\subsubsection{Projections}\label{sec: lightmediatorprojections}

\begin{figure}[t]
\centering
  \includegraphics[width=0.47\linewidth]{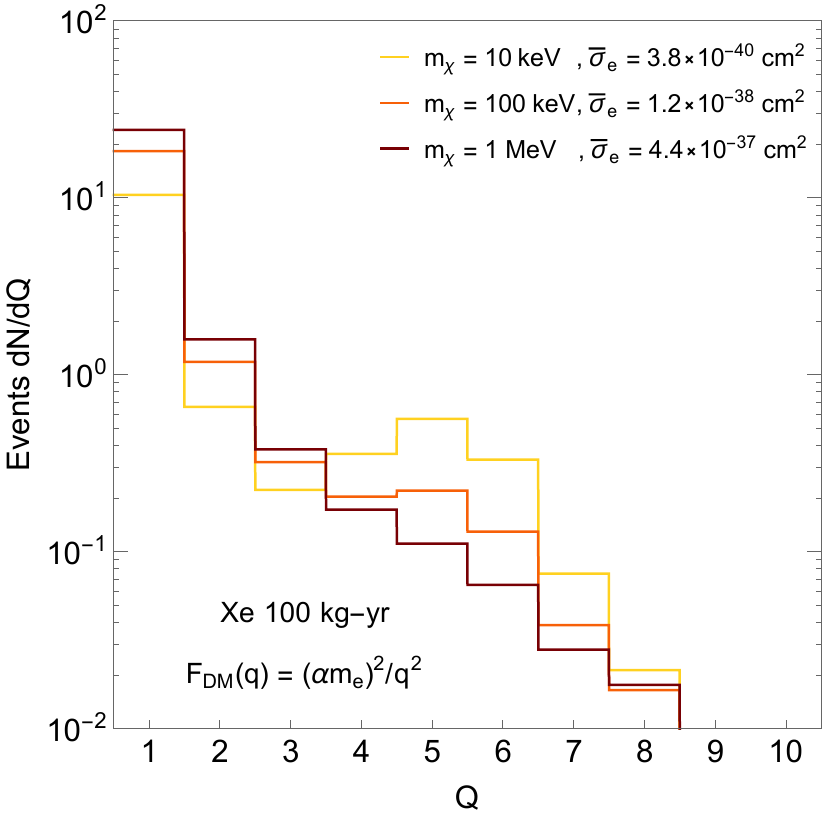}
  \includegraphics[width=0.47\linewidth]{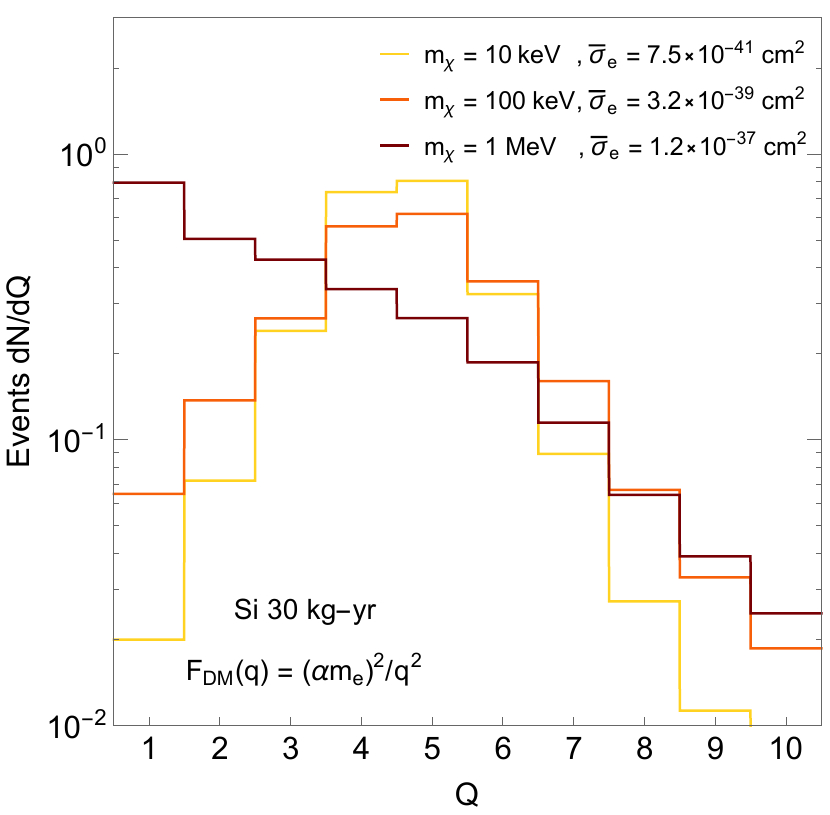}
\caption{Electron recoil spectra produced by the SRDM flux scattering in a xenon target (left, 100~kg-year exposure) and a silicon target (right, 30~kg-yr exposure) for DM coupled to an  ultralight dark photon. DM masses are as indicated, and the cross sections are chosen to be at around the 90\% CL sensitivity projections shown in \cref{fig: lightmediatorprojections}.}
\label{fig: xe100kgyr_and_oscura_signals}
\end{figure}

We present here projections for upcoming and proposed DM detectors using noble-liquid and silicon targets. We ignore backgrounds and show curves in the $\overline\sigma_e$ versus $m_\chi$ plane that correspond to the 90\% CL sensitivity for a background-free experiment. We show both the projections for the SRDM flux component as well as for the halo DM component. For proposed experiments with a xenon target (such as LBECA~\cite{Bernstein:2020cpc}), we show projections in \cref{fig: lightmediatorprojections} (left) for a 10 kg-yr and a 100 kg-yr exposure; for the former, we assume a threshold of 2 electrons, while for the latter, we show curves corresponding to thresholds of 2, 4, and 6 electrons. We see that such xenon detectors can probe part of the freeze-in line for sub-MeV DM, in particular for $m_\chi\lesssim$ 200~keV.  In \cref{fig: xe100kgyr_and_oscura_signals} (left), we show the electron spectra induced by the SRDM flux in a xenon target for three DM masses ($m_\chi=10$~keV, $m_\chi=100$~keV, and $m_\chi=1$~MeV), for a 100 kg-yr exposure, and for cross sections at the value that corresponds to the 90\% CL sensitivity for each DM mass. 

For upcoming and proposed experiments with silicon targets, we show projections in \cref{fig: lightmediatorprojections} (right) for a 100~g-yr, 1~kg-yr, and a 30~kg-yr exposure; for the 100~g-yr and 1~kg-yr exposures, we show curves for a threshold of 2 electrons, while for the 30~kg-yr exposure, we show curves corresponding to thresholds of 2, 3, and 4 electrons.  The different exposures correspond to the exposure goals of SENSEI~\cite{Tiffenberg:2017aac,Crisler:2018gci,SENSEI:2019ibb,SENSEI:2020dpa,SENSEI:2023zdf}, DAMIC-M~\cite{DAMIC-M:2023gxo}, and Oscura~\cite{Oscura:2022vmi} experiments.  We show both the projections for the SRDM flux component as well as for the halo DM component. 
It is noteworthy that with an exposure of 30 kg-yr, the entire freeze-in benchmark can be probed, even for sub-MeV DM masses. 
We also see that the sensitivities between the various chosen thresholds is very similar. This can be understood by considering the expected number of events versus the electron recoil spectrum generated by the SRDM flux. In \cref{fig: xe100kgyr_and_oscura_signals} (right), we show the electron spectra of SRDM in a silicon target for three DM masses ($m_\chi=10$~keV, $m_\chi=100$~keV, and $m_\chi=1$~MeV), for a 30~kg-year exposure, and for cross sections at the value that corresponds to the 90\% CL sensitivity for each DM mass. For lower DM masses, the SRDM flux has a high-velocity component, which is able to excite the plasmon in silicon and yield a peak in the $Q=4-6$ bins~\cite{Essig:2024ebk} (we discuss this further in \cref{app: silicon form factors}). This makes the SRDM sensitivity projections for DM with masses below 1~MeV largely insensitive to whether the threshold is 2, 3, or 4 electrons. For DM masses near 1~MeV, the resulting SRDM flux is less boosted, so that  the plasmon does not dominate the spectrum and the electron recoil spectra peak at low $Q$.

\subsection{Direct-detection results for dark matter coupled to a heavy dark photon}
\begin{figure}[t]
    \centering
    \includegraphics[width=0.6\textwidth]{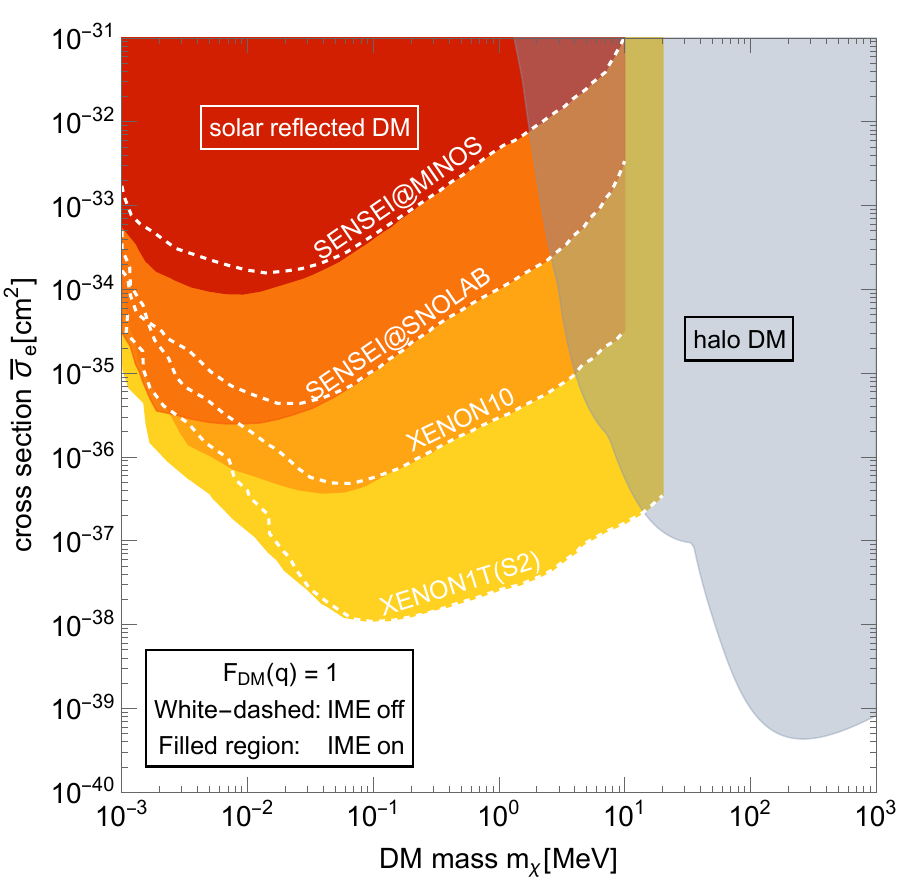}
    \caption{Solar reflection exclusion limits (90\% CL) from direct-detection experiments for DM coupled to a heavy dark photon. The excluded regions shaded in solar colors are from SENSEI@MINOS~\cite{SENSEI:2020dpa}, SENSEI@SNOLAB~\cite{SENSEI:2023zdf}, XENON10~\cite{Angle:2011th, Essig:2012yx, Essig:2017kqs} and XENON1T (S2-only)~\cite{XENON:2019gfn}. For halo DM, the excluded region is the combined limit from SENSEI@MINOS~\cite{SENSEI:2020dpa}, XENON10~\cite{Angle:2011th, Essig:2012yx, Essig:2017kqs} and XENON1T (S2-only)~\cite{XENON:2019gfn}. The white dashed lines indicate the limits when in-medium effects (``IME'') in the Sun are ignored.}
    \label{fig: heavymediatorbound}
\end{figure}

In this section, we focus on DM coupled to a heavy dark photon.  We present existing bounds from the SRDM flux as well as projections for upcoming and proposed experiments with xenon and silicon targets.  

\subsubsection{Exclusion limits}

Using the same form factors and detector response models as in \cref{sec: lightmediatorbound}, we show the SRDM constraints (at 90\% CL) for DM coupled to a heavy dark photon in \cref{fig: heavymediatorbound}. Excluded regions from direct-detection experiments for SRDM are shaded in solar colors. The contour for halo limits (shaded in gray) is composed by constraints from SENSEI@MINOS~\cite{SENSEI:2020dpa}, XENON10~\cite{Angle:2011th, Essig:2012yx, Essig:2017kqs} and XENON1T (S2-only)~\cite{XENON:2019gfn}. We see that the SRDM component allows direct-detection experiments to exclude vast region of sub-MeV DM parameter space. We compare the bounds with and without the inclusion of in-medium effects in the Sun (the latter are shown with white dashed lines). In-medium effects in the Sun are important especially for $m_\chi\lesssim$ 100~keV, where they allow for more energetic reflected DM particles (see \cref{sec: results-flux-heavy}).

\subsubsection{Projections}

\begin{figure*}[t]
    \centering
    \includegraphics[width=\textwidth]{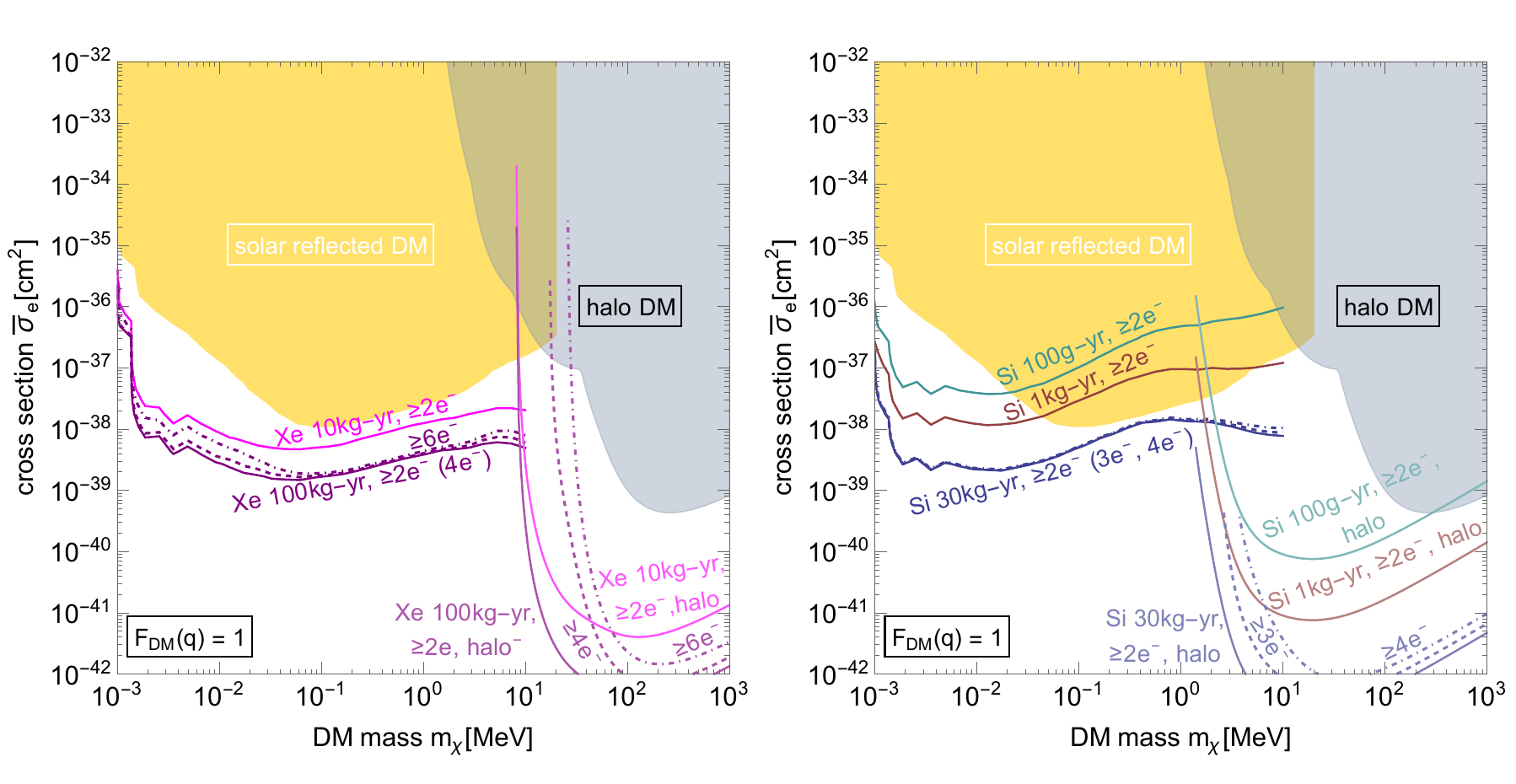}
    \caption{
    Projections for the solar-reflected DM flux for upcoming and proposed DM detectors with xenon (left) or silicon (right) targets, for DM coupled to a heavy dark photon. The left plot shows SRDM projections for xenon detectors for a 10~kg-yr exposure and a threshold of 2~electrons (magenta curve), 100~kg-yr and a threshold of 2~electrons, 4~electrons, or 6~electrons (solid, dashed, dot-dashed purple curves). The right plot shows SRDM projections for silicon detectors with a 100~g-yr exposure and a threshold of 2 electrons (green curve), 1 kg-yr and 2 electrons (red curve), 30 kg-yr and 2, 3, or 4 electrons (solid, dashed, dot-dashed blue curves). The yellow region is the combined SRDM limits from \cref{fig: heavymediatorbound} and the other lines and shaded regions are as in \cref{fig: heavymediatorbound}.
    }
    \label{fig: heavymediatorprojections}
\end{figure*}

\begin{figure}[t]
\centering
  \includegraphics[width=0.47\linewidth]{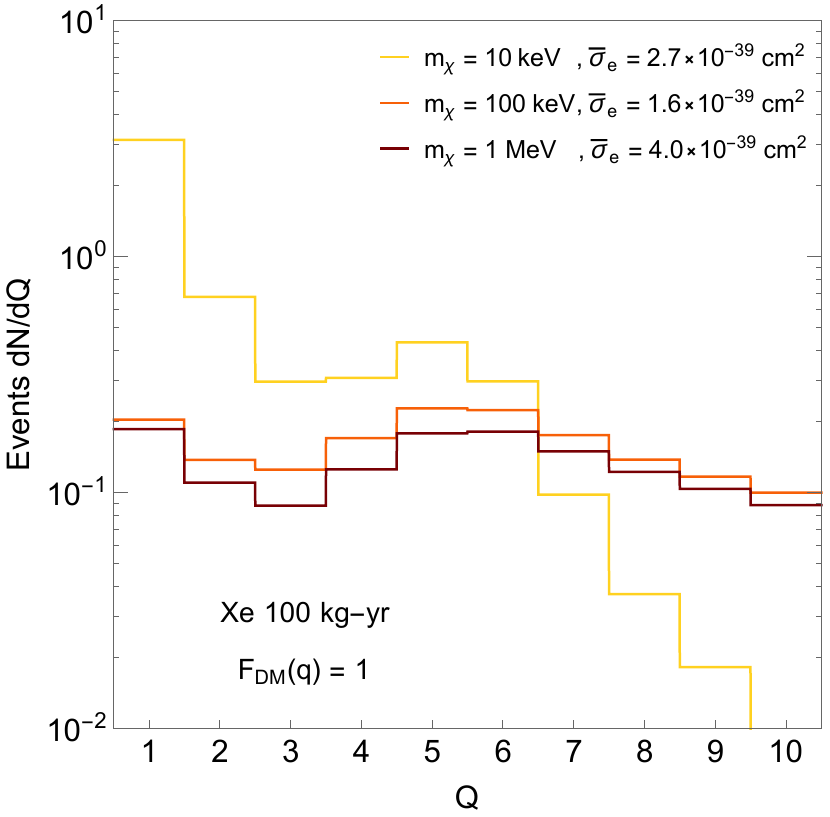}
  \includegraphics[width=0.47\linewidth]{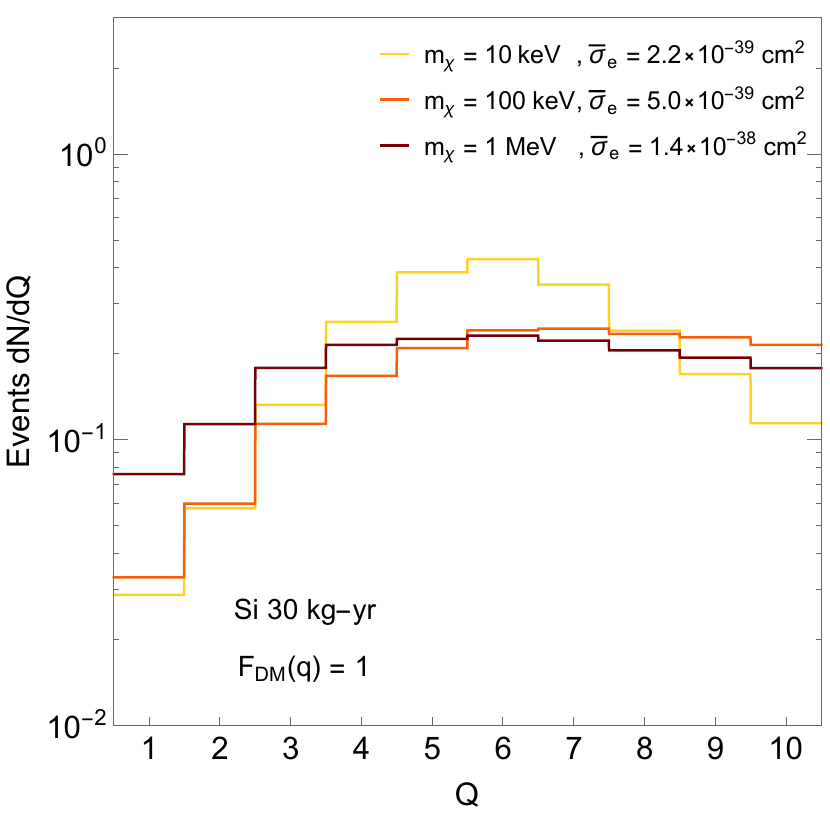}
\caption{Electron recoil spectra produced by the SRDM flux scattering in a xenon target (left, 100~kg-year exposure) and a silicon target (right, 30~kg-yr exposure) for DM coupled to a heavy dark photon. DM masses are as indicated, and the cross sections are chosen to be at around the 90\% CL sensitivity projections shown in \cref{fig: heavymediatorprojections}. 
}
\label{fig: xe100kgyr_and_oscura_signals_heavymediator}
\end{figure}

We present SRDM projections, along with the projections using the halo DM component, for upcoming and proposed detectors with xenon and silicon targets for DM with a heavy dark-photon mediator in \cref{fig: heavymediatorprojections}.  We show sensitivities for the same assumptions about the exposures and thresholds as in \cref{sec: lightmediatorprojections}. In \cref{fig: xe100kgyr_and_oscura_signals_heavymediator}, we show the electron spectra of SRDM in a xenon target for a 100~kg-year exposure (left) and in a silicon target for a 30~kg-year exposure (right) for three DM masses ($m_\chi=10$~keV, $m_\chi=100$~keV, and $m_\chi=1$~MeV), and for cross sections at the value that corresponds to the 90\% CL sensitivity for each DM mass.

\subsection{Annual signal modulation of solar-reflected dark matter}\label{sec: modsignalrate}

\begin{figure}[t]
    \centering
    \includegraphics[width=0.6\textwidth]{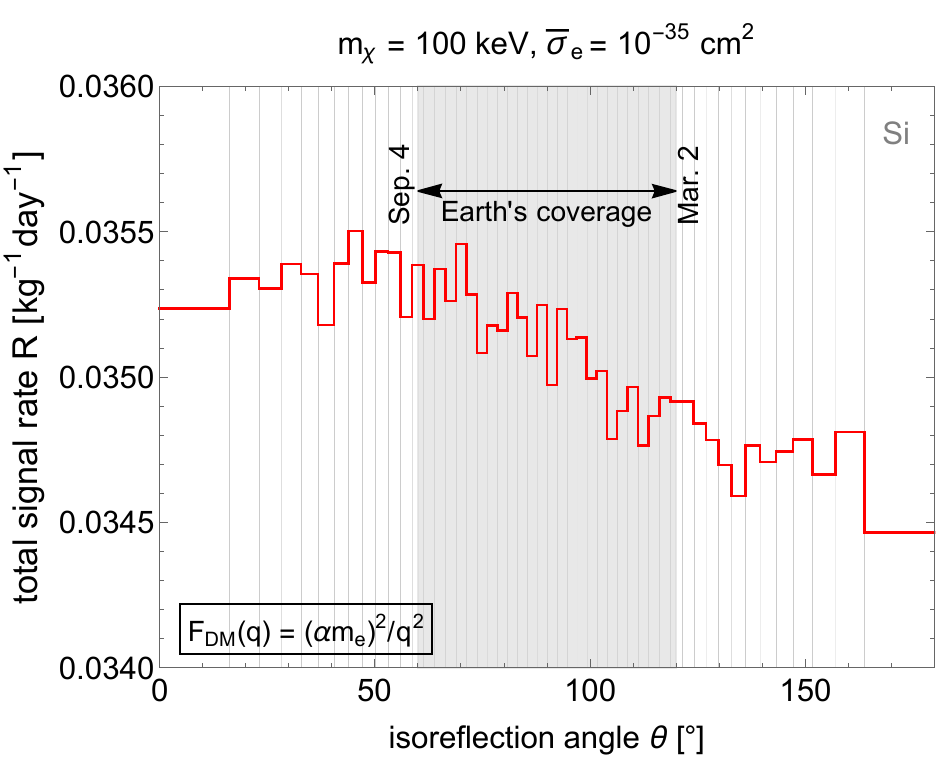}
    \caption{The predicted SRDM signal rate $R$ in a silicon detector as a function of isoreflection angle $\theta$, obtained by MC simulations with 50 isoreflection rings. We fix the Earth-Sun distance to be 1~AU, so the modulation shown here includes only the modulation due to the anisotropy of the incoming particles to the Sun.}
    \label{fig: anisosignal}
\end{figure}
The annual modulation of the SRDM signal comes from two sources. The first is the anisotropy of the solar reflection as a consequence of the anisotropy of the in-coming DM particles to the Sun. The second is the orbital modulation from the Earth's distance to the Sun, which varies by $\sim$3\%  over the year. The annual modulation of the signal observed on Earth is the combination of these two effects.

To study the annual modulation signal, we consider a DM mass of $m_\chi=100$~keV and a cross section of $\overline\sigma_e=10^{-35}\;\text{cm}^2$. We simulate the solar reflection with 50 isoreflection rings and calculate the signal rate of SRDM in a silicon detector. To focus on the signal modulation of the solar reflection due to the anisotropy of the incoming particles to the Sun, we fix the Earth's distance to the Sun at 1~AU. The result is shown in \cref{fig: anisosignal}. The signal modulation from the anisotropy of the flux leads to a deviation from the mean value that is less than 1\%.  This is largely due to the high-energy tail of the flux spectrum varying very little over the year (see ~\cref{fig: anisoflux_and_diffflux})). In this case, the most significant origin of the annual modulation of the signal rate is due to the modulation of the orbital distance; the amplitude of the annual modulation is $\sim$3\%, maximized (minimized) at the perihelion (aphelion) on $\sim$Jan.~3 ($\sim$Jul.~4), which coincidentally is almost exactly out of phase with the annual modulation of the halo DM signal. Although small, it may still be possible to search for the annual modulation of the SRDM signal rate with future detectors. 

\subsection{Directional detection of solar-reflected dark matter}
As the Earth moves in the galaxy, the velocity distribution of the halo DM at the Earth is anisotropic, and there is a ``DM wind'' coming from the opposite direction of the Earth's motion in the galactic rest frame. If the target material used in a DM direct-detection experiment has a directional response, then one can anticipate a daily modulation of the DM scattering rate, since a target fixed to the Earth rest frame will rotate with respect to the DM wind. The directional dependence can make it easier to distinguish potential DM signals from backgrounds. Recent proposals for target materials with a directional response for low-mass DM include two-dimensional targets such as graphene~\cite{Hochberg:2016ntt, Catena:2023qkj, Catena:2023awl}, anisotropic Dirac materials~\cite{Coskuner:2019odd, Geilhufe:2019ndy}, and anisotropic crystals~\cite{Griffin:2018bjn}. Proposals also exist for using isotropic superconductors to look for an anisotropic distribution of final state excitations from the anisotropy of incoming DM~\cite{Hochberg:2021ymx}.

\begin{figure}[t]
    \centering
    \includegraphics[width=0.6\textwidth]{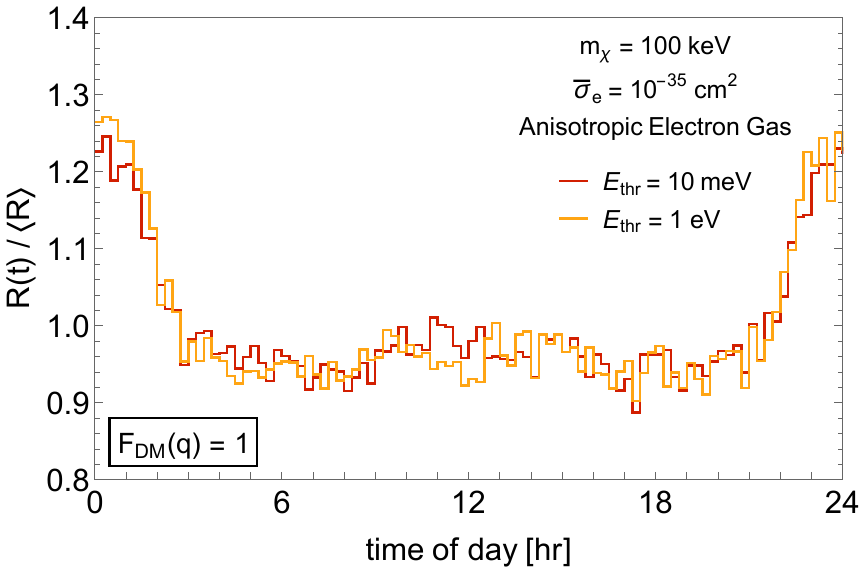}
    \caption{The SRDM flux is expected to induce a daily modulation in the signal rate when scattering off a target material with an anisotropic response.  To illustrate the daily modulation, the figure shows the time-dependent signal rate (normalized to the average daily rate $\left<R\right>$) for the SRDM scattering off an anisotropic electron gas for DM coupled to a heavy dark photon for two different thresholds, 10~meV (red) and 1~eV (orange) for $m_\chi=100$~keV and $\overline{\sigma}_e=10^{-35}$~cm$^2$. }
    \label{fig: aniso_electron_gas}
\end{figure}

Since the SRDM flux comes from essentially a point source (the Sun), one could use target materials with a directional response to search for the SRDM flux. To illustrate this, we model the target material as an anisotropic electron gas~\cite{Boyd:2022tcn}. The anisotropic electron gas has different effective masses $m_{x,y,z}$ along the spacial directions $\{x,y,z\}$. We assume the same fiducial parameters for the target as in Fig.~2 of~\cite{Boyd:2022tcn} and include a minimum energy transfer to mimic the detector threshold. We use the convention that the $z$-direction of the target is aligned with the Sun at time $t=0$ of the day. The resulting daily modulation of the SRDM signal is shown in \cref{fig: aniso_electron_gas} for a DM mass $m_\chi=100$~keV, a cross section $\overline\sigma_e=10^{-35}\;\text{cm}^2$, for DM coupled to a heavy dark photon, and for the energy threshold of 10~meV and 1~eV. The daily modulation of SRDM for both energy thresholds are $\sim$20\%. It shows that for sub-MeV DM, the SRDM flux is able to induce a strong modulation signal in anisotropic materials with $\sim$eV-scale threshold, for which there will be no signal from the halo DM component.

%%%%%%%%%%%%%%%%%%%%%%%%%%%%%%%%%%%%%%%%%%%%%%%%%%%%%%%%%%%%%%%
\section{Conclusions}\label{sec: conclusions}
Halo DM particles can be attracted to the Sun by its gravitational pull, scatter off particles in the hot solar plasma, and get boosted to higher velocities. This ``solar-reflected DM'' (SRDM) component dramatically extends the sensitivity of terrestrial DM detectors to lower DM masses, and in many cases provides the leading bounds on sub-MeV DM. In this paper, we studied the properties of the SRDM flux and its phenomenology in direct-detection experiments, focusing especially on DM coupled to a dark photon. The results in this work are based on the MC tool \texttt{DaMaSCUS-SUN}~\cite{Emken2021}, which simulates individual DM particles passing through the Sun, generates the SRDM flux, and evaluates the SRDM signals in various detectors. Along with this paper, we make available the SRDM fluxes for a wide range of DM masses and cross sections.

The main model considered in this paper is fermionic DM coupled to a dark photon that is kinetically mixed with the Standard Model photon. We focus on ultralight dark-photon mediators, but also revisit heavy dark-photon mediators.  We include the in-medium screening effects of dark photons in the solar plasma, and we include the latest silicon form factors that include collective effects (such as the plasmon) to calculate the scattering in direct-detection experiments.

For DM coupled to an ultralight dark photon, which is in general only weakly constrained below 1~MeV, we find strong exclusions limits on sub-MeV DM using current data from XENON10, XENON1T, and SENSEI, surpassing the SN1987A bound by more than one order of magnitude.  We also project sensitivities of upcoming and proposed detectors, including the silicon detectors SENSEI, DAMIC-M, and Oscura (with anticipated exposures of 100~g-yr, 1~kg-yr, and 30~kg-yr, respectively) and xenon detectors with no low-energy ionization background (with exposures of 10~kg-yr and 100~kg-yr). Remarkably and notably, an Oscura-like detector can probe the entire freeze-in benchmark model, with the SRDM flux allowing the benchmark to be probed for sub-MeV DM masses.  This statement holds even if the background-free threshold for Oscura is ``only'' 4~electron-hole pairs, since the electron recoil spectrum induced by the SRDM flux peaks near 4 to 6 electron-hole pairs, at least for DM masses below $\sim$100~keV.

DM coupled to a heavy dark photon is in general more strongly constrained for sub-MeV masses due to the bounds on the number of relativistic degrees of freedom~\cite{Boehm:2013jpa}. Nevertheless, we update the SRDM bounds that were previously considered in e.g.~\cite{Emken:2021lgc, An:2021qdl} to include in-medium effects on the dark photon interaction in the solar plasma as well as the improved silicon form factors for boosted particles.  The in-medium effects turn out to be important for DM masses below $\sim$100~keV as their inclusion provides more high-velocity particles and leads to stronger constraints.

We also investigate the modulation in the scattering rate induced by the SRDM flux. The annual modulation of the SRDM signal comes from two sources. One is the fluctuation of the distance between the Earth and the Sun, another is the anisotropy of the SRDM flux due to the anisotropic initial condition of DM falling into the Sun. The annual modulation of the SRDM signal has a different pattern compared with the annual modulation of the halo DM signal, which could potentially allow the SRDM signal to be distinguished from backgrounds or from the halo DM signal. We also considered the directional detection of the SRDM flux, pointing out that it may provide a powerful handle for distinguishing an SRDM signal from backgrounds.

In conclusion, the SRDM flux provides a powerful tool to extend the reach of direct-detection experiments to sub-MeV DM.
%%%%%%%%%%%%%%%%%%%%%%%%%%%%%%%%%%%%%%%%%%%%%%%%%%%%%%%%%%%%%%%
\acknowledgments
We thank Haipeng An, Haoming Nie, Maxim Pospelov, and Josef Pradler for useful discussions about their previous solar-reflected dark matter results.  H.X.~thanks Haipeng An and Haoming Nie for their hospitality hosting him at Tsinghua University.  T.E.~thanks the Theoretical Subatomic Physics group at Chalmers University of Technology for its hospitality. We also thank Ryan Plestid and Aman Singal for useful discussions about the impact of plasmons in silicon on the expected spectra from the scattering of boosted dark matter, and Aman Singal for providing the silicon form factors. We thank Christian Forssen for discussions about Monte Carlo methods.
T.E.~was supported by the Knut and Alice Wallenberg Foundation (PI, Jan Conrad).
R.E.~acknowledges support from the US Department of Energy under Grant DE-SC0009854, 
from the Heising-Simons Foundation under Grant No.~79921, from the Simons Foundation under the Simons Investigator in Physics Award~623940, from the Binational Science Foundation under Grant No.\ 2020220, from Stony Brook IACS Seed Grant, and from Fermilab subcontract 664693 for the DoE DMNI award for Oscura. 
H.X.~is supported in part by DoE Grant DE-SC0009854 and by the Simons Investigator in Physics Award 623940. 

\appendix

\section{The photon self-energy in the solar medium}\label{app: inmedium}
To account for in-medium effects for DM scattering in the solar plasma through a dark-photon mediator, the photon self-energy needs to be calculated. The leading-order contribution of the self-energy comes from the one-loop diagram: 
\begin{figure*}[h]
    \centering
    \includegraphics[width=0.2\textwidth]{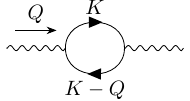}
\end{figure*}
\\In a medium with finite temperature, we use the thermal field theory presented in~\cite{lebellac_1996} to calculate the diagram above. The expression is
\begin{equation}\label{poltensorexpress}
    \begin{split}
    \Pi^{\mu\nu}(Q)=4e^2\int\frac{\dd^3\vec{k}}{(2\pi)^3}\frac{1}{2E_k}(n_e(E_k)+n_{\bar{e}}(E_k))\frac{Q\cdot K(K^\mu Q^\nu+Q^\mu K^\nu)-Q^2K^\mu K^\nu-(Q\cdot K)^2g^{\mu\nu}}{(Q\cdot K)^2-Q^4/4}\,,
    \end{split}
\end{equation}
where $Q=(q^0,\vec{q})$ is the four-momentum of the photon, $K=(E_k,\vec{k})$ is the on-shell four-momentum with $E_k=\sqrt{m_e^2+k^2}$, and $n_e(E)=(e^{(E- \mu_e)/T}+1)^{-1}$ and $n_{\bar{e}}(E)=(e^{(E+ \mu_e)/T}+1)^{-1}$ is the occupation number of electrons and positrons, respectively, for a particle with a spin given by the Fermi-Dirac distribution in a medium with temperature $T$ and chemical potential $\mu_e$.

Based on the general formula in \cref{poltensorexpress}, we next follow~\cite{DeRocco:2022rze} for calculating the photon self-energy in the solar medium. The solar medium can be treated as a dilute, non-relativistic gas, so we use a Maxwell-Boltzmann distribution for the occupation number instead of a Fermi-Dirac distribution. The electron velocity follows a Maxwell-Boltzmann velocity distribution
\begin{equation}\label{mbdis}
    n_e=n_e\int \dd^3\vec{v}\,p(v),\quad p(v)=\frac{1}{(2\pi\sigma^2)^{3/2}}e^{-v^2/(2\sigma^2)}\,,
\end{equation}
where $\sigma=\sqrt{T/m_e}$. We need to relate $p(v)$ with the distribution in momentum space $n_e(E_k)$ in \cref{poltensorexpress}. In momentum space, we have
\begin{equation}
    n_e=2\int\frac{\dd^3\vec{k}}{(2\pi)^3}n_e(E_k)=\frac{2m_e^3}{(2\pi)^3}\int \dd^3\vec{v}\,n_e(E_{k=m_ev})\,,
\end{equation}
where the factor of 2 is the electron's spin degeneracy. Comparing this relation with \cref{mbdis}, we obtain
\begin{equation}\label{mvdis}
    n_e(E_k)=(2\pi)^3\frac{n_e}{2m_e^3}p(k/m_e)\,.
\end{equation}
For non-relativistic scattering, the scattering through the longitudinal polarization dominates (see e.g.~\cite{Hochberg:2015fth}), and so we only need the longitudinal self-energy $\Pi_L(Q)$ here. It is related to the self-energy tensor by $\Pi_L(Q)=\frac{Q^2}{q^2}\Pi^{00}(q)$ as can be seen from \cref{poltensor}. Now we change the variable from momentum $\vec{k}$ to $\vec{v}=\vec{k}/m_e$ in \cref{poltensorexpress} and insert in the Maxwell-Boltzmann velocity distribution into \cref{mvdis}, to obtain 
\begin{equation}\label{pilq}
    \begin{split}
        \Pi_L(Q)\simeq&\frac{Q^2}{q^2}\frac{e^2 n_e}{m_e}\int \dd^3\vec{v}\,p(v)\frac{2q^0(q^0-\vec{v}\cdot \vec{q})-Q^2-(q^0-\vec{v}\cdot \vec{q})^2}{(q^0-\vec{v}\cdot \vec{q})^2-Q^4/4m_e^2}\\
        =&\frac{Q^2}{q^2}\omega_p^2\int \dd^3\vec{v}\,p(v)\frac{q^2-(\vec{v}\cdot \vec{q})^2}{(q^0-\vec{v}\cdot \vec{q})^2-Q^4/4m_e^2}\,,
    \end{split}
\end{equation}
where $\omega_p^2\equiv e^2n_e/m_e$ is the plasma frequency. It can be further simplified realizing that $(\vec{v}\cdot \vec{q})\ll q$ so $(\vec{v}\cdot \vec{q})^2$ can be neglected in the numerator. Also, the integrand in \cref{pilq} only depends on the projection of $\vec{v}$ onto $\vec{q}$, so we can define $v_q=\vec{v}\cdot \vec{q}/q$ and integrate out the other two components of $\vec{v}$ that are perpendicular to $\vec{q}$,
\begin{equation}
    \Pi_L(q)\simeq\omega_p^2Q^2\int \dd v_q\,p_q(v_q)\frac{1}{(q^0-qv_q)^2-Q^4/4m_e^2}\,,
\end{equation}
where $p_q(v_q)=\frac{1}{\sqrt{2\pi\sigma^2}}e^{-v_q^2/(2\sigma^2)}$ is the Maxwell-Boltzmann velocity distribution for $v_q$.  
We then have
\begin{equation}
    \begin{split}
        \Pi_L(q)\simeq\omega_p^2 m_e\int \dd v_q\frac{1}{\sqrt{2\pi\sigma^2}}e^{-v_q^2/(2\sigma^2)}\left(\frac{1}{qv_q-(q^0+Q^2/2m_e)}-\frac{1}{qv_q-(q^0-Q^2/2m_e)}\right)\,.
    \end{split}
\end{equation}
We can express the integral in terms of the plasma dispersion function~\cite{fitzpatrick2022plasma}
\begin{equation}
    Z(z)=\frac{1}{\sqrt{\pi}}\int_{-\infty}^\infty \dd x\frac{e^{-x^2}}{x-z}=\sqrt{\pi}e^{-z^2}(i-Erfi(z))\,.
\end{equation}
\begin{figure}[t]
    \centering
    \includegraphics[width=0.5\textwidth]{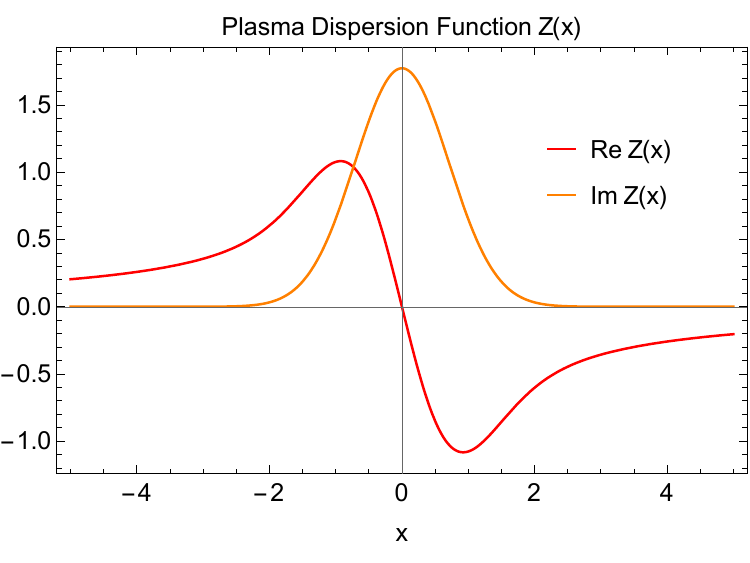}
    \caption{Plots of real (blue) and imaginary (orange) part of the plasma dispersion function $Z$ evaluated at real axis.}
    \label{fig: IME-Zfunc}
\end{figure}
A plot of $Z(x)$ is shown in \cref{fig: IME-Zfunc}. Defining $\xi(Q)=\frac{q^0}{\sqrt{2}\sigma q}$ and $\delta(Q)=\frac{-Q^2}{2\sqrt{2}qm_e\sigma}$, we have
\begin{equation}\label{polfinalresult}
    \Pi_L(Q)\simeq\omega_p^2\frac{m_e}{q^0}\xi(Z(\xi-\delta)-Z(\xi+\delta))\,.
\end{equation}
This is the electron's contribution to the longitudinal photon self-energy. For the contribution from the other charged components in the solar plasma with number density $n_i$, mass $m_i$, and charge number $Z_i$, the calculation is very similar: we can simply use \cref{polfinalresult} with the appropriate number density and mass, and after multiplying $Z_i^2$ in the prefactor.

\section{Details on the Metropolis algorithm for the solar-reflected flux}
\label{app: simulation details}

In \cref{eq: PDF q cos theta}, we showed the probability distribution of the momentum transfer~$q$ and the angle~$\theta$ between~$\mathbf{v}_\chi$ and~$\mathbf{q}$ (or, more precisely, $\cos\theta)$.
It was given as
\begin{align*}
        f_i(q,\cos\theta) &= \frac{1}{\Omega_i(r,v)} \frac{\dd^2\Omega_i}{\dd q \dd\cos\theta}\, .
\end{align*}
Every time a DM~particle scatters inside the Sun, we need to sample~$(q,\cos\theta)$ from this joint distribution.
For light mediator and in-medium effects, this challenges standard sampling methods such as inverse-transform sampling or rejection sampling.
For one, the evaluation of the PDF is computationally expensive due to the normalization constant that requires numerical integration of complex functions. 
Secondly, the probability mass of the distribution might be concentrated in a tiny sub-volume of the~$(q,\cos\theta)$ domain.

Both challenges can be overcome by using the Metropolis algorithm~\cite{Metropolis1953}.
The Metropolis algorithm is a Markov Chain Monte Carlo (MCMC) method to obtain a sequence of samples from multidimensional distributions~$f(\mathbf{x})$ that cannot simply be sampled directly.
It generates a sequence $\{\mathbf{x}_1,\mathbf{x}_2,\mathbf{x}_3,...\}$, where the longer this sequence becomes the closer it follows the actual target distribution.
In doing so, it is not required to evaluate the actual normalized PDF~$f_i(q,\cos\theta)$, but instead it suffices to evaluate a function~$g(q,\cos\theta) $ that is proportional to the PDF, i.e.~$f_i(q,\cos\theta) \sim g(q,\cos\theta) $.
In our case, this means that the computationally challenging normalization constant may be ignored, and we only evaluate the differential scattering rate,
\begin{align}
     g(q,\cos\theta)  = \frac{\dd^2\Omega_i}{\dd q \ \dd\cos\theta}\, ,
\end{align}
to obtain a~$(q,\cos\theta)$ sample.

\begin{figure*}[t]
    \centering
    \includegraphics[width=0.49\textwidth]{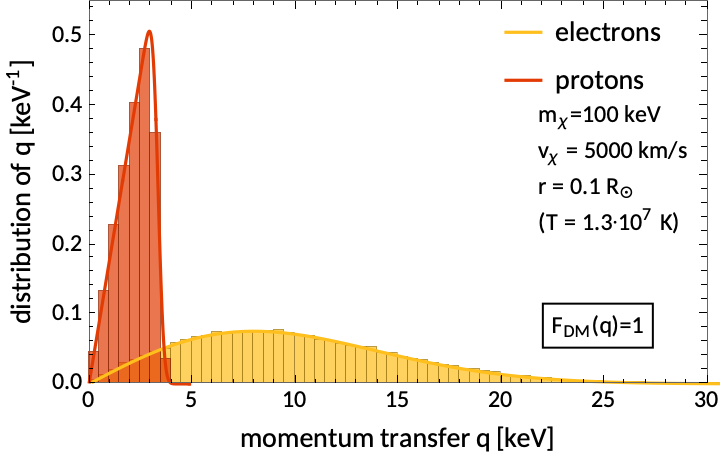}
    \includegraphics[width=0.49\textwidth]{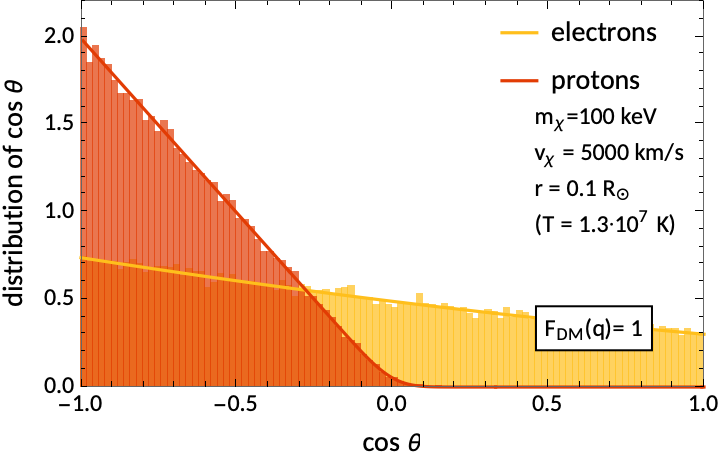}\\
    \includegraphics[width=0.49\textwidth]{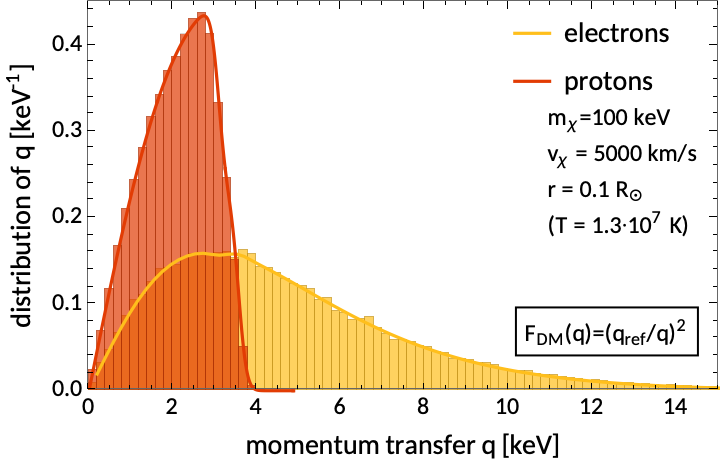}
    \includegraphics[width=0.49\textwidth]{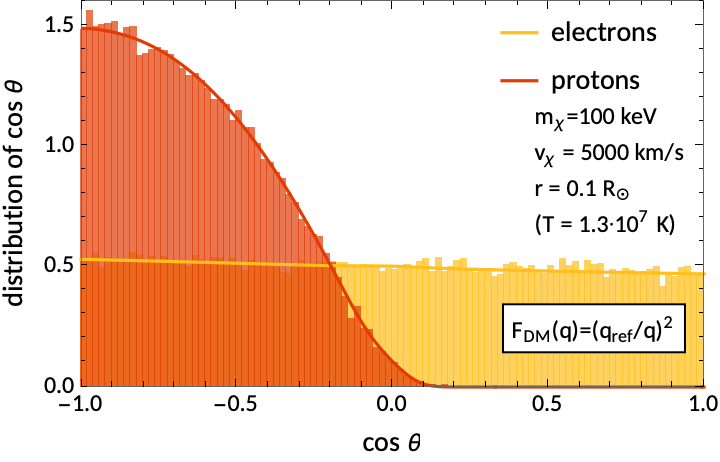}
    \caption{Random samples of the momentum transfer~$q$ (left) and angle~$\cos\theta$ (right) for scatterings with electrons (yellow) and protons (red) with in-medium effects. The first (second) row depicts contact (long range) interactions. We used the Metropolis algorithm to sample 50000~$(q,\cos\theta)$ pairs (seen in the normalized histograms) and compare them to the marginal PDFs of~$q$ and~$\cos\theta$ respectively (solid lines) .}
    \label{fig: metropolis sampling validation}
\end{figure*}

With regards to the second challenge, the Metropolis algorithm automatically identifies the region of the domain where the probability mass is concentrated.
This typically happens during the ``burn-in'' phase, as described below.
Lastly, the problem of correlated samples within the sequence, typical for this kind of algorithm, does not arise in our simulations as we only need a single~$(q,\cos\theta)$ sample for each scattering, for which we start a new Markov chain. 

The algorithm starts by picking a random point~$\mathbf{x} = (q,\cos\theta)$ from the domain~$(0,q_\mathrm{max})\times(-1,1)$.
Therefore, our sequence or Markov chain starts as~$\{\mathbf{x}\}$.
Based on this point, we sample a new `candidate' point $\mathbf{y} = (q^\prime,\cos\theta^\prime)$ from a chosen `proposal distribution'.
Here, we use a two-dimensional Gaussian distribution centered at~$\mathbf{x}$ with standard deviations~$\boldsymbol{\sigma} = (\sigma_q, \sigma_{\cos\theta})$.\footnote{In our simulations, we use~$\sigma_q = \frac{\sigma_\mathrm{max}}{5}$ and~$\sigma_{\cos\theta} = 0.2$.}
The candidate point~$\mathbf{y}$ is accepted as an addition to the sequence (or Markov chain) with a probability of~$\alpha$ given by
\begin{align}
    \alpha = \mathrm{min}\left(1, \frac{g(\mathbf{y})}{g(\mathbf{x})}\right)\, .
\end{align}
If the new value is accepted, we add~$\mathbf{y}$ to the Markov chain, otherwise we add the previous value~$\mathbf{x}$.
Then, we repeat the procedure by sampling a new candidate.

The procedure resembles a random walk through the distribution's domain, where the walker takes a small step or remains standing depending on the distribution.
If the next candidate point is more probable than the current point, the algorithm always accepts it.
Otherwise, there is a chance of rejecting the candidate and remain at the current point.
After an initial phase of getting settled (the `burn-in' phase) the resulting sequence of~$\mathbf{x}$ values closely follow the target distribution~$f_i(q,\cos\theta)$.
In our simulation, we use a burn-in phase of 1000 steps.
In other words, we disregard the first 1000 members of the Markov chain and use the next point as the sample of the momentum transfer~$q$ and angle~$\cos\theta$.

We have performed extensive tests, where we compared histograms of Metropolis-generated samples with the marginalized target~PDFs~$f(q)$ and~$f(\cos\theta)$, which verified that the resulting samples' distribution approximates~$f_i(q,\cos\theta)$ with high precision.
One example of such a verification is shown in \cref{fig: metropolis sampling validation}.

\section{Comparison with other xenon form factors}\label{app: xenon form factors}

\begin{figure}[t]
    \centering
    \includegraphics[width=\textwidth]{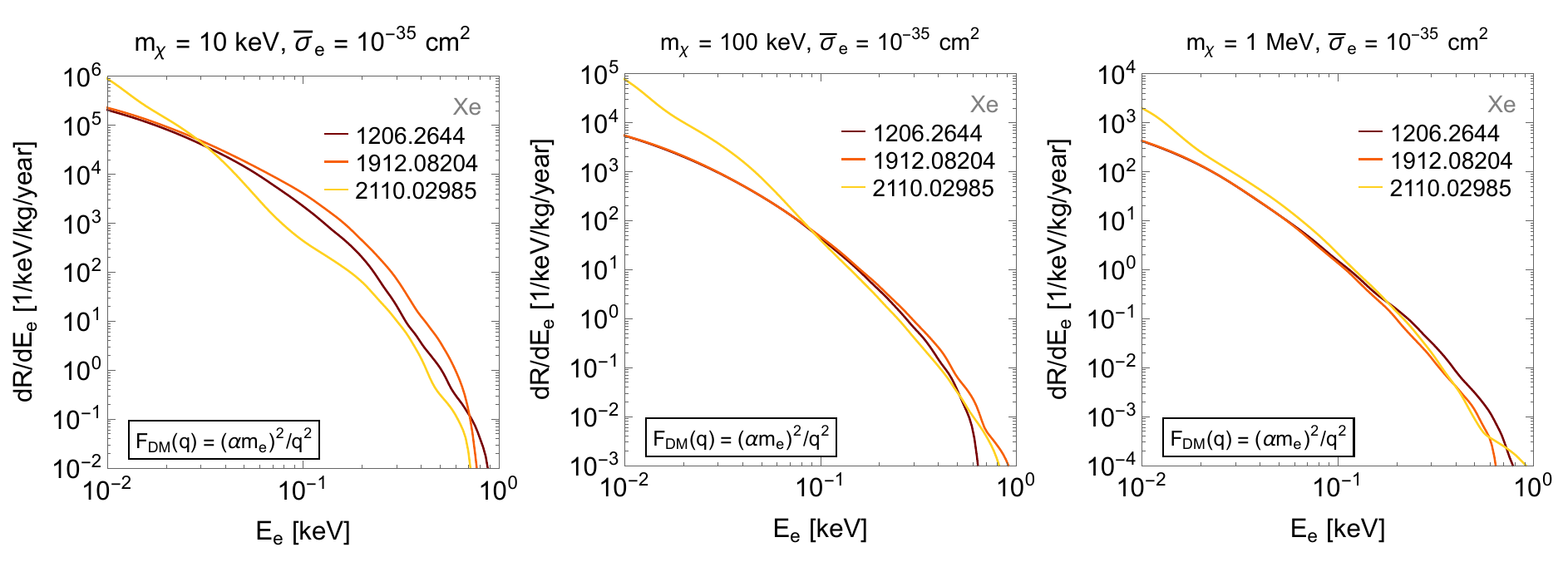}
    \caption{The energy spectra in a xenon target induced by the SRDM flux for DM coupled to an ultralight dark photon, evaluated with three different calculations of the xenon form factor (as indicated in the plot legends), for three choices of the DM mass.}
    \label{fig: 3XeFFcompare}
\end{figure}
\begin{figure}[t]
    \centering
    \includegraphics[width=0.6\textwidth]{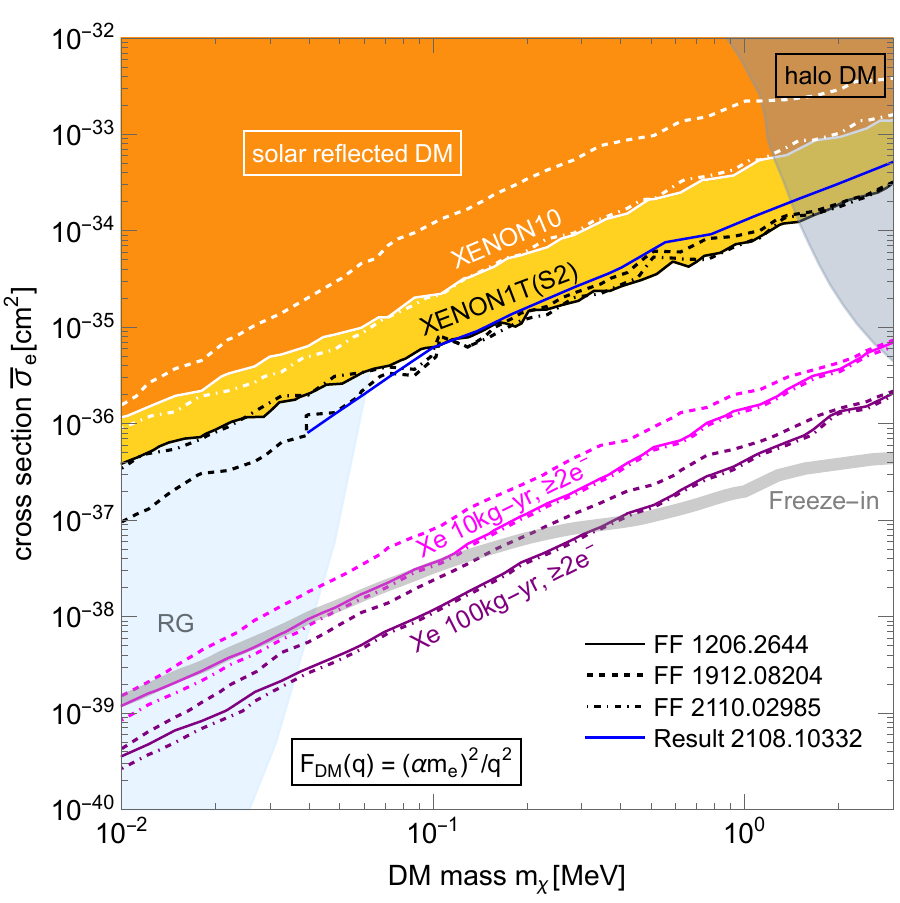}
    \caption{Solar reflection exclusion limits and projections (90\% CL) from xenon-based direct-detection experiments for DM coupled to an ultralight dark photon. Using the simulations of the SRDM flux presented in this paper, we compare the constraints and projections for three different publicly-available xenon form factors~\cite{Essig:2012yx,Catena:2019gfa, Hamaide:2021hlp}. We also compare the XENON1T (S2-only) bound derived in this paper with the one derived in Ref.~\cite{An:2021qdl}, which uses an independent simulation of the SRDM flux, an independent calculation of the xenon form factor (using the FAC package~\cite{Gu:2008}), and an independent analysis of the bound from the XENON1T data.  The excluded regions shaded in solar colors are from XENON10~\cite{Angle:2011th, Essig:2012yx, Essig:2017kqs}, and XENON1T (S2-only)~\cite{XENON:2019gfn}. The region shaded in light blue on the left is the red giants cooling constraint~\cite{Vogel:2013raa}. For halo DM, the excluded region is the combined limit from SENSEI@SNOLAB~\cite{SENSEI:2023zdf} and SENSEI@MINOS~\cite{SENSEI:2020dpa}. The freeze-in benchmark is plotted in gray~\cite{Essig:2011nj,Chu:2011be,Essig:2015cda,Dvorkin:2019zdi}.}
    \label{fig: 3XeFF_bounds_and_projections}
\end{figure}
For the constraints and projections from xenon-based detectors, we use the form factor calculated in~\cite{Essig:2012yx}, which is also used in subsequent works, e.g.~\cite{Essig:2017kqs, XENON:2019gfn}. There are, however, other calculations for the xenon form factors that are publicly available, e.g.~\cite{Catena:2019gfa, Hamaide:2021hlp}. We compare the energy spectra induced by the SRDM flux for these various form factors in \cref{fig: 3XeFFcompare}. In addition, we compare the impact of the different form factors on the SRDM constraints from XENON10 and XENON1T (S2-only) as well as on the projections for a background-free 10~kg-yr and 100~kg-yr exposure experiment in \cref{fig: 3XeFF_bounds_and_projections}. The only difference between the various lines labeled by ``FF" is the choice of the xenon form factor that affects the direct-detection sensitivity, but SRDM flux simulations are the same. We see that the constraints and projections agree generally within a factor of a few. Moreover, in \cref{fig: 3XeFF_bounds_and_projections} we compare our XENON1T (S2-only) constraint (black solid curve) for SRDM with the result from Fig.~10 of~\cite{An:2021qdl} (blue solid curve with the label ``Result 2108.10332"). Ref.~\cite{An:2021qdl} features an independent simulation of the SRDM flux, a different xenon form factor (based on the FAC code~\cite{Gu:2008} and also used in, e.g.,~\cite{Ibe:2017yqa,Essig:2019xkx}), and an independent analysis of the detector response. We have confirmed that the SRDM fluxes from both groups agree well with each other, and believe that the difference (less than a factor of 3) most likely arises due to the use of a different xenon form factor and/or a different treatment of the XENON1T data. 
Overall though, the agreement is good.

\begin{figure*}[t]
    \centering
    \includegraphics[width=\textwidth]{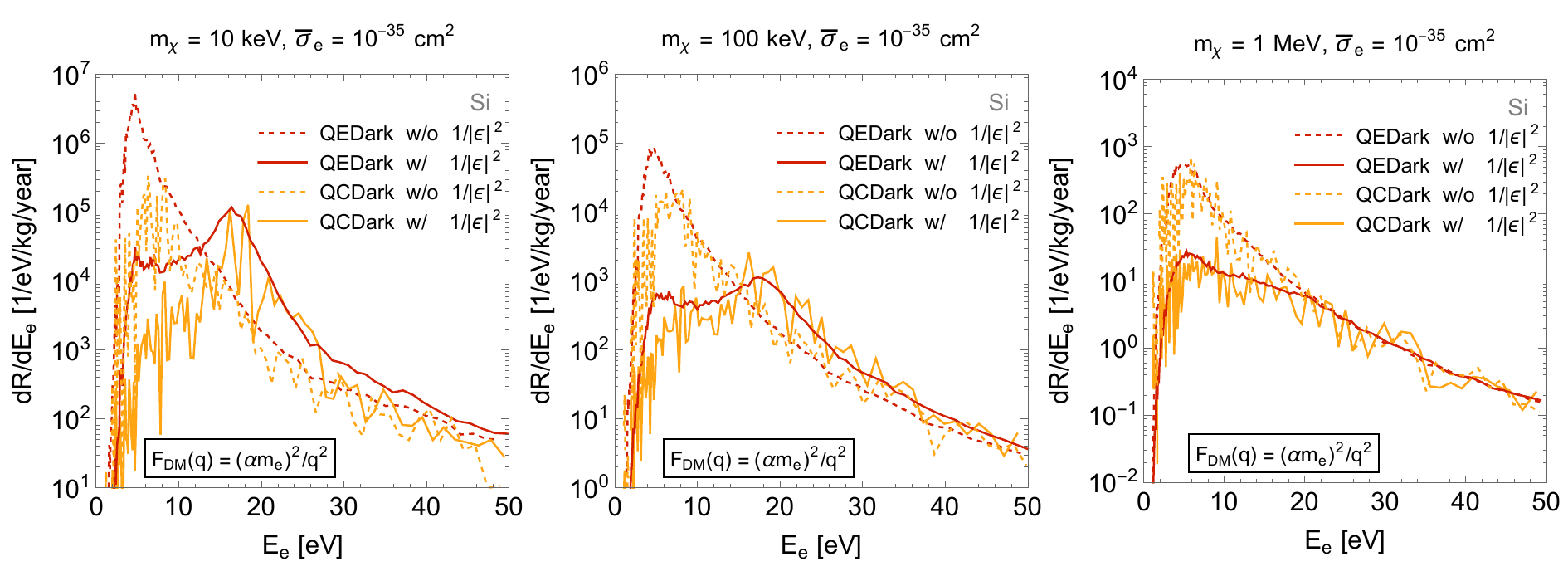}
    \caption{Electron recoil spectrum in a silicon target induced by the SRDM flux for DM coupled with an ultralight dark photon for DM masses of 10~keV (left), 100~keV (middle) and 1~MeV (right), respectively. We compare the rates when using \texttt{QCDark} with collective effects (written as ``w/ $1/|\epsilon|^2$'', where $\epsilon$ is the dielectric function) in the form of the Lindhard dielectric function (which is the default calculation that we use for the main results in the paper) to the rates when using \texttt{QCDark} without collective effects (written as ``w/o $1/|\epsilon|^2$''), as well as when using \texttt{QEDark} with, and without, collective effects.}
    \label{fig: 4SiFFcompare}
\end{figure*}

\begin{figure*}[t]
    \centering
        \includegraphics[width=\textwidth]{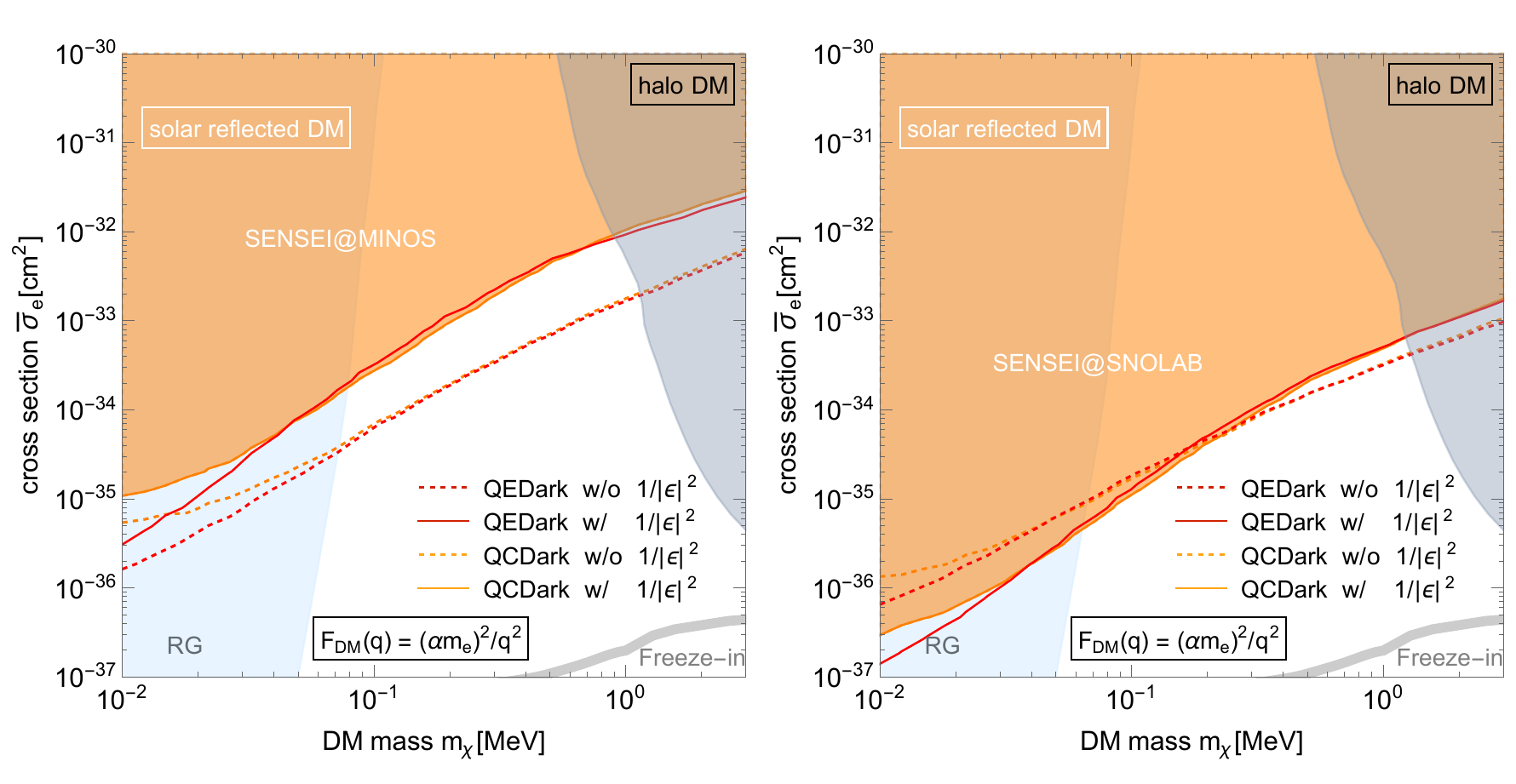}
    \caption{SRDM constraints from SENSEI@MINOS (left) and SENSEI@SNOLAB (right), for DM coupled to an ultralight dark photon. Electron recoil spectrum in a silicon target induced by the SRDM flux for DM coupled with an ultralight dark photon.  We compare the rates when using \texttt{QCDark} with collective effects (written as ``w/ $1/|\epsilon|^2$'', where $\epsilon$ is the dielectric function) in the form of the Lindhard dielectric function (which is the default calculation that we use for the main results in the paper) to the rates when using \texttt{QCDark} without collective effects (written as ``w/o $1/|\epsilon|^2$''), as well as when using \texttt{QEDark} with, and without, collective effects.  The other shaded regions and the freeze-in line is as in \cref{fig: lightmediatorbound}.}
    \label{fig: 4SiFF_lightmediator_bound}
\end{figure*}

\section{Comparison with other silicon form factors}\label{app: silicon form factors}

For the constraints and projections from silicon-based detectors, we use \texttt{QCDark}~\cite{Dreyer:2023ovn,QCDark} and correct it with the Lindhard dielectric function~\cite{lindhard1954properties, Knapen:2021run, Hochberg:2021pkt} to account for in-medium screening and plasmon excitations, as discussed in~\cref{sec: crystal excitation} and in~\cite{Essig:2024ebk}. We show in this section the impact of the collective effects and of using a different silicon form factor on the SRDM-induced electron recoil spectrum and the resulting constraints and projections. In particular, \cref{fig: 4SiFFcompare} shows the induced SRDM electron-recoil spectrum for three different DM masses when using \texttt{QEDark}~\cite{Essig:2015cda,QEDark} instead of \texttt{QCDark} and with and without the inclusion of collective effects (written as ``w/ $1/|\epsilon|^2$'' or ``w/o $1/|\epsilon|^2$'', respectively, where $\epsilon$ is the dielectric function). Compared with the spectra when using \texttt{QEDark}, the spectra when using \texttt{QCDark} have a lower rate for $E_e\lesssim15$~eV and a similar rate for higher $E_e$. Additionally, the inclusion of the collective effects modelled through the dielectric function attenuates the peak of the spectra by about two orders of magnitude due to in-medium screening, but also produces a peak in the spectrum for energies in the range $\left[15, 20\right]$~eV due to the excitation of the plasmon. The plasmon resonance is stronger for lighter DM masses, because the Sun can boost lighter DM to much higher velocities, allowing for the excitation of the plasmon~\cite{Essig:2024ebk}. For example, while the plasmon peak is still clearly visible for $m_\chi\simeq100$~keV, it is not visible anymore for $m_\chi\simeq1$~MeV. The peak in the spectrum between 15 to 20~eV corresponds to about 4 to 6 electrons, which also explains why the sensitivity to the SRDM flux does not degrade much when changing the energy threshold from 2 electrons to 4 electrons in \cref{fig: lightmediatorprojections}. This is also important since backgrounds in current detectors peak towards lower energies~\cite{SENSEI:2020dpa, DAMIC-M:2023gxo,SENSEI:2023zdf}.

\cref{fig: 4SiFF_lightmediator_bound} shows the impact of choosing different form factors and of the collective effects on the constraints and projections. For SRDM with an ultralight dark photon, using $\texttt{QCDark}$ results in slightly weaker constraints than using $\texttt{QEDark}$ for $m_\chi\lesssim30$ keV, which can be understood from \cref{fig: 4SiFFcompare}. The inclusion of collective effects with $1/\left|\epsilon\right|^2$ can result in either stronger or weaker bounds, depending on which charge bins are being considered.  For example, for the SENSEI@MINOS limit, we consider the charge bins $Q=1-6$~electrons, while for SENSEI@SNOLAB, we consider the charge bins $Q=4-10$~electrons.

%\normalem
\bibliographystyle{JHEP}
\bibliography{main.bbl}

\end{document}